\documentclass[pra,nofootinbib,floats,aps,twocolumn,tightenlines,superscriptaddress]{revtex4-1}

\usepackage{graphicx}
\usepackage{bbold}
\usepackage{mathtools}
\usepackage{hyperref}
\usepackage{commath}
\usepackage{bm}
\usepackage{amsfonts,amsmath,amssymb,amsthm}
\usepackage[usenames,dvipsnames]{xcolor}
\usepackage{subfigure}
\usepackage{dsfont}

\usepackage{simplewick}

\usepackage{qcircuit}
\usepackage[normalem]{ulem}

\DeclareMathOperator{\Tr}{Tr}

\renewcommand*\d[2][]{%
	\mathrm{d}%
	\ifx\relax#1\relax\else
	\rule{-0.02em}{1.5ex}^{#1}\rule{0.08em}{0ex}\!
	\fi
	#2\,
}
\newcommand{\ket}[1]{| {#1} \rangle}
\newcommand{\bra}[1]{\langle {#1} |}
\newcommand{\tr}{\text{Tr}}

\newcommand{\ii}{\mathrm{i}}

\newcommand{\ad}[1]{\hat{a}_{\bm{#1}}^\dagger}

\newcommand{\rhoa}{\hat{\rho}_\textsc{a}}
\newcommand{\rhob}{\hat{\rho}_\textsc{b}}
\newcommand{\rhoab}{\hat{\rho}_\textsc{ab}}

\newcommand{\ya}[1]{\hat Y_{\textsc{a}#1}}
\newcommand{\yb}[1]{\hat Y_{\textsc{b}#1}}

\newcommand*{\QED}{\hfill\ensuremath{\square}}
\newcommand{\ta}[1]{t_{\textsc{a}_#1}}
\newcommand{\tb}[1]{t_{\textsc{b}_#1}}

\newcommand{\ca}[1]{\hat y_{\textsc{a}#1}^+}
\newcommand{\cb}[1]{\hat y_{\textsc{a}#1}^+}
\newcommand{\sa}[1]{\hat y_{\textsc{a}#1}^-}
\renewcommand{\sb}[1]{\hat y_{\textsc{a}#1}^-}
\newcommand{\cnu}[1]{\hat y_{\nu#1}^+}
\newcommand{\snu}[1]{\hat y_{\nu#1}^-}
\newcommand{\oa}{\Omega_\textsc{a}}
\newcommand{\ob}{\Omega_\textsc{b}}

\newcommand{\integral}[3]{\int_{#2}^{#3} \!\! \mathrm{d} #1 \,}
\newcommand{\comm}[2]{[{#1},{#2}]}

\newcommand{\ketbra}[2]{\left| {#1}\vphantom{#2} \right\rangle\!\left\langle {#2}\vphantom{#1} \right|}

\newcommand{\UA}[1]{\hat{U}_{\textsc{a}_{#1}}}
\newcommand{\UB}[1]{\hat{U}_{\textsc{b}_{#1}}}

\newcommand{\XA}[1]{\hat{X}_{\textsc{a}_{#1}}}
\newcommand{\XB}[1]{\hat{X}_{\textsc{b}_{#1}}}

\newcommand{\mA}[1]{\hat{m}_{\textsc{a}_{#1}}}
\newcommand{\mB}[1]{\hat{m}_{\textsc{b}_{#1}}}

\newcommand{\id}{\mathds{1}}

\newcommand{\nn}{\nonumber\\}

\begin{document}

\title{General no-go theorem for entanglement extraction}
	
\author{Petar Simidzija}	\affiliation{Department of Applied Mathematics, University of Waterloo, Waterloo, Ontario, N2L 3G1, Canada}
\affiliation{Institute for Quantum Computing, University of Waterloo, Waterloo, Ontario, N2L 3G1, Canada}

\author{Robert H. Jonsson}
\affiliation{QMATH, Department of Mathematical  Sciences,  University  of  Copenhagen,
Universitetsparken  5,  2100  Copenhagen,  Denmark}
	
\author{Eduardo Mart\'in-Mart\'inez}
\affiliation{Department of Applied Mathematics, University of Waterloo, Waterloo, Ontario, N2L 3G1, Canada}
\affiliation{Institute for Quantum Computing, University of Waterloo, Waterloo, Ontario, N2L 3G1, Canada}
\affiliation{Perimeter Institute for Theoretical Physics, Waterloo, Ontario N2L 2Y5, Canada}

\begin{abstract}

We study under what circumstances a separable bipartite system A-B can or cannot become entangled through local interactions with a bi-local entangled source $\text{S}_1$-$\text{S}_2$. We obtain constraints on the general forms of the interaction Hamiltonians coupling A with $\text{S}_1$ and B with $\text{S}_2$ necessary for A and B to become entangled. We are able to generalize and provide  non-perturbative insight on several previous no-go theorems of entanglement harvesting from quantum fields using these general results. We also discuss the role of communication in the process of entanglement extraction, establishing a distinction between genuine entanglement extraction and communication-assisted entanglement generation.

\end{abstract}
	
\maketitle

\section{Introduction}
\label{sec:intro}

One of the most striking differences between classical and quantum theories is the existence of entanglement in the latter. The implications of this fact are enormous from both practical as well as fundamental viewpoints. The practical importance of entanglement is perhaps most evident in the advantages that it seems to offer quantum computers over their classical counterparts~\cite{Horodecki2009}. On the more fundamental side, for example, the presence of entanglement in the vacuum state of a quantum field~\cite{Summers1985,Summers1987} is a key feature in conjectured solutions to the black hole information loss problem~\cite{Preskill1992,Susskind1993,Almheiri2013,Braunstein2013}. 

Given the relevance  of entanglement, both as a resource for quantum information processing and as a tool in fundamental studies, it is important that we better understand its dynamical nature. In this context, we analyze the ability of bipartite systems to extract entanglement from sources in which entanglement already exists naturally. In particular, we concern ourselves with the ability of a separable target system A-B to extract entanglement from a bipartite source system $\text{S}=\text{S}_1$-$\text{S}_2$ only through bilocal interactions A-$\text{S}_1$, B-$\text{S}_2$. More concretely, we ask the question: Given the unitary operator $\hat U$ which describes the interaction between target and source, what conditions must $\hat U$ obey in order for A and B to become entangled?

We find conditions that are necessary in order to achieve entanglement extraction with a particular class of generic target-source interactions. Specifically, we consider interactions where the unitary $\hat U$ is the exponential of a Schmidt rank 1 operator.
Based on the observation that such interactions give rise to an entanglement breaking channel from the source to the target, we show that, in particular, entanglement extraction requires more than two such interactions (i.e., more than just one per target). We give necessary conditions for combinations of three or four such couplings to achieve entanglement extraction. In these cases where A and B \emph{do} become entangled, we discuss the origin of this entanglement. That is, we establish a distinction between the genuine extraction of pre-existing entanglement from S, and the generation of entanglement between A and B due to indirect communication via the source S.

These main results apply generally to the framework of entanglement extraction independent of the physical manifestation of the target and source systems. In this paper, we establish their significance in particular for \emph{entanglement harvesting} from relativistic quantum fields: By a direct application of our main result, we generalize all previously known no-go results for entanglement harvesting and provide a simple, unified explanation for why they hold.

In so called entanglement harvesting, which is a special case of entanglement extraction, the targets A and B are a pair of first-quantized systems, e.g., modeled by Unruh-DeWitt (UDW) particle detectors~\cite{DeWitt1979}, and the source S is a relativistic quantum field. The pioneering works on entanglement harvesting were by Valentini~\cite{Valentini1991}, and later Reznik \textit{et. al.}~\cite{Reznik2003,Reznik2005}, and they showed that it is possible for particle detectors A and B to become entangled through local interactions with the field vacuum, \emph{even if the detectors are spacelike separated}. Since spacelike separated detectors cannot communicate with one another, this provided a simple operational proof of the fundamental fact that the field vacuum contains entanglement with respect to local modes.

Following these initial works, the entanglement harvesting protocol has been studied in much further detail. For instance it has been shown that it is possible (albeit more difficult) to harvest entanglement from thermal states in timelike~\cite{Braun2002,Braun2005} and spacelike \cite{Brown2013a} separation. It has also been shown that entanglement can be harvested from coherent~\cite{Simidzija2017c} scalar field states, as well as from the electromagnetic field vacuum using fully featured hydrogen-like atoms~\cite{Pozas2016}. The sensitivities of the protocol to the properties~\cite{Pozas2015} and trajectories~\cite{Salton2015} of the detectors, boundary conditions of the field~\cite{Brown2014}, nature of the detector-field couplings~\cite{Sachs2017}, as well as the geometry~\cite{Steeg2009} and topology~\cite{Martinez2016a} of the background spacetime have also been investigated. 

Besides their fundamental significance, the above studies are important in determining the optimal conditions for an experimental realization of an entanglement harvesting protocol. On a positive note, it has been suggested that such a protocol may be within reach using current atomic and superconducting setups~\cite{Olson2011,Olson2012,Sabin2012}, and in principle could provide a constant supply of Bell pairs which could later be used for quantum information purposes in \textit{entanglement farming} protocols~\cite{Martinez2013a}. 
However, many aspects, in particular with respect to  potential  implementations, still need to be explored; one important question is the energetic cost of entanglement harvesting, 
which could be particularly high in a low number of spatial dimensions, as recently addressed in Ref.~\cite{Beny2017}.

Whereas much of the previous literature focused on perturbative analyses of entanglement harvesting, the interaction between particle detectors and relativistic fields can be analyzed non-perturbatively in certain particular setups. For example, significant work has been done to develop tools that allow for the non-perturbative study of harmonic oscillator detectors in diverse contexts (see, e.g. \cite{BlokSYuin,Bruschi2013,Brown2013}). On the other hand, for finite-dimensional detectors, non-perturbative time-evolution can be computed when the detector's Hamiltonian is completely degenerate (i.e., all detector states have identical energies~\cite{Braun2002,Braun2005,Landulfo2016}), or when the detector interacts with the field at one instant in time (i.e, via a Dirac-$\delta$ coupling)~\cite{Hotta2008,Hotta2009}. In particular, using these approaches, the following no-go entanglement harvesting theorems were proved: i) Perturbatively, it is not possible to harvest spacelike vacuum entanglement with zero-gap detectors \cite{Pozas2017}, and ii) non-perturbatively it is not possible to harvest any kind of entanglement (timelike, lightlike, or spacelike) from a coherent field state using single $\delta$-coupled detectors~\cite{Simidzija2017c}.

In this paper, using our general result on the inability of target systems A and B to become entangled with a source S, we will non-perturbatively explain the results i) and ii) stated above using a single mathematical formalism, while at the same time generalizing them to hold for \emph{any} field state. 

This paper is organized by generality: We start in Sec.~\ref{sec:general_ent_extraction} by outlining the general setup of separable targets A and B that attempt to become entangled by interacting with a source S. In Sec.~\ref{sec:ent_breaking_channels} we prove that if A and B each interact with S through a single ``simple-generated" unitary (one that is the exponential of a Schmidt rank 1 operator), then they cannot become entangled. In Sec.~\ref{sec:3interactions}, we show that if one of the targets couples to S through \emph{two} simple-generated interactions, then A and B can become entangled under certain conditions. Then, in Sec.~\ref{sec:null_result_ent_harv} we particularize these results to the setup of UDW detectors interacting with a relativistic quantum field, which allows us to generalize two previous no-go entanglement harvesting theorems regarding i) degenerate detectors and ii) single Dirac-$\delta$ coupled detectors. In Sec.~\ref{sec:simple_setup_delta_harvesting} we show how two $\delta$-couplings for one of the detectors could lead to entanglement harvesting. Surprisingly, and in stark contrast with previous perturbative studies, we find that for detector-field couplings in the non-perturbative regime, an increase in coupling strength leads to a \emph{decrease} in the amount of harvested entanglement. Natural units of $\hbar=c=1$ are used throughout.

\section{Entanglement extraction with simple-generated interactions}\label{sec:2}

This section presents the central result of the article. Whereas we later focus on its  implications for entanglement harvesting from a relativistic field, the result applies to the  wider, general framework of entanglement extraction from a source system to two target systems. Therefore, we begin with a brief review of the general setup.

Our main result refers to the interactions between the source and target systems having a particular simple shape, that is, that the unitary describing the interaction is given by the exponential of a simple tensor product operator. We show that a single simple-generated interaction of this kind gives rise to an entanglement breaking channel from the source to the target, such that entanglement extraction with only one interaction per target is impossible. Instead, when using simple-generated interactions, at least one target has to couple twice, i.e., at least three interactions in total are required. We discuss further necessary conditions on this minimal scenario and illustrate them in terms of toy models.

\subsection{Entanglement extraction from a source}
\label{sec:general_ent_extraction}

We begin by reviewing the framework of entanglement extraction. The general idea is that two parties A and B want to entangle their local quantum systems (the targets) by extracting correlations that are originally contained in a third quantum system (the source, S). 

For instance, the source might be a spatially extended quantum system such as a quantum many-body system or a quantum field. A and B couple to separate parts of S, the latter being in a state that contains entanglement between spatially separated degrees of freedom. Examples of such states include the ground states of interacting lattice theories or the vacuum state of a quantum field.

The total Hilbert space of the two targets and the source is the tensor product 
\begin{align}
    \mathcal{H}=\mathcal{H}_\textsc{a}\otimes\mathcal{H}_\textsc{s}\otimes\mathcal{H}_\textsc{b},
\end{align}
of the Hilbert space of the source $\mathcal{H}_\textsc{s}$, and of the two targets $\mathcal{H}_\textsc{a}$ and $\mathcal{H}_\textsc{b}$. The latter two we here assume to be finite-dimensional,  while we allow $\mathcal{H}_\textsc{s}$ to be of any (finite or infinite) dimension.
Initially, before any interactions take place, the three subsystems start out in a product state
\begin{align}
    \hat \rho_0=\hat\rho_{0,\textsc{a}}\otimes\hat\rho_{0,\textsc{s}}\otimes\hat\rho_{0,\textsc{b}},
\end{align}
and the free dynamics of the system are generated by the sum of the three free Hamiltonians
\begin{align}
    \hat H_0=\hat H_\textsc{A}+\hat H_\textsc{S}+\hat H_\textsc{B},
\end{align}
where $\hat H_\textsc{s}$ is shorthand for $\id_\textsc{a}\otimes \hat H_\textsc{s}\otimes \id_\textsc{b}$, etc.

We suppose that the two parties each only have access to a limited part of the source, e.g., only to a certain region of spacetime when the source is a relativistic quantum field. Their goal is to swap entanglement  that is present in the state $\hat \rho_{0,\textsc{s}}$ between their respective regions of access, onto their local target systems. 

To achieve this the two targets A and B locally couple to the source through time-dependent interaction Hamiltonians  $\hat H_{\textsc{i,a}}(t)$ and $\hat H_{\textsc{i,b}}(t)$ (shorthand for $\hat H_{\textsc{i,a}}(t)\otimes\id_\textsc{b}$, etc.). However, no direct interaction Hamiltonian between the two targets is allowed. Also, we do not assume any classical communication between the parties.

The interaction between A, B and S is then given by a unitary operator $\hat U$ acting on the Hilbert space $\mathcal{H}$. Calculating this evolution operator is in general difficult. A common approach, in particular in the context of perturbation theory, is to write $\hat U$ as a Dyson expansion. In the interaction picture, this is
\begin{align}\label{eq:U_dyson}
    \hat U
    =\id-\ii\integral{t}{}{} \hat H_\textsc{i}(t) -\integral{t}{}{}\integral{t'}{}t \hat H_\textsc{i}(t) \hat H_\textsc{i}(t') +... \quad ,
\end{align}
where $\hat H_\textsc{i}(t):=\hat H_{\textsc{i,a}}(t)+\hat H_{\textsc{i,b}}(t)$.
An alternative method of expressing $\hat U$ is via the Magnus expansion~\cite{blanes_magnus_2009}, which expresses the operator as
\begin{align}\label{eq:U_magnus}
    &\hat U =\exp \left(\sum_{n=1}^\infty \hat \Omega_n \right),
\end{align}
where the lowest order terms read
\begin{align}
    &\hat \Omega_1= -\ii \integral{t}{-\infty}\infty \hat H_\textsc{i}(t), \nn
    &\hat \Omega_2 = -\frac12 \integral{t}{-\infty}\infty \integral{t'}{-\infty}t \comm{\hat H_\textsc{i}(t)}{\hat H_\textsc{i}(t')},
\end{align}
and the higher-order terms, which are obtained recursively, contain commutators of the commutators of the interaction Hamiltonian $\hat H_\textsc{i}(t)$ at increasing orders. 

The results and discussion of this section apply to any system where the unitary $\hat U$ describing one coupling between a target (A or B) and the source S has the property of being the exponential of an operator with Schmidt rank 1. That is, $\hat U$ is generated by a Hamiltonian which is a simple tensor product of two observables:
\begin{align}\label{eq:r1generated}
    \hat U=\exp\left(-\ii\, \hat m\otimes \hat X\right),
\end{align}
with $\hat m$ acting on one of the targets and $\hat X$ acting on the source system. In the following we will refer to such interactions either as Schmidt rank 1 generated or just as \emph{simple generated}.

\subsection{Single simple-generated interaction yields entanglement breaking channel}\label{sec:ent_breaking_channels}

We will now show that no entanglement can be extracted from the source if both targets each couple to the source system with only one simple-generated interaction. Instead, entanglement extraction can only be achieved if at least one of the targets is coupled to S through at least two simple interactions.

The reason for this is that when a target (say A) and the source, which are initially in a product state, interact via a single simple-generated unitary $\hat U_\textsc{a}$, the map from the source's initial state to the final state of the target,
\begin{align}\label{eq:ent_breaking_channel}
    \hat \rho_{0,\textsc{s}}\mapsto\hat \rho_\textsc{a} = \Tr_\textsc{s} \left[\hat U_\textsc{a} \left(\hat \rho_{0,\textsc{a}}\otimes\hat \rho_{0,\textsc{s}}\right)\hat U_\textsc{a}^\dagger\right],
\end{align}
is an entanglement breaking quantum channel.

entanglement breaking channels are characterized by the property that when they receive only a part of a larger system as input, which may be entangled with other degrees of freedom, then the output of the channel is always in a separable state with the rest of the larger system. That is, any entanglement between the input and the environment is broken and the output is not entangled with the environment anymore~\cite{horodecki_general_2003}.

To see why channels of the form \eqref{eq:ent_breaking_channel} are entanglement breaking, let us recast the simple-generated unitary $\hat U_\textsc{a}$ in the form of a controlled unitary:
\begin{align}\label{eq:controlled_unitary}
    \hat U_\textsc{a}&=\exp\left(-\ii \, \hat m\otimes \hat X\right) \notag \\
    &= \sum_k \exp\left( -\ii x_k \hat m \right) \otimes\ketbra{x_k}{x_k}.
\end{align}
Here we are assuming that the self-adjoint operator $\hat X$, acting on the source system, has the  discrete spectral decomposition $\hat X=\sum_k x_k \ketbra{x_k}{x_k}$ with $x_k\in\mathbb{R}$. This is true by the spectral theorem if $\hat X$ is a compact (and hence bounded) operator. The more general case of a non-bounded $\hat X$ --- which is indeed the case if $\hat X$ is a smeared field operator acting on the Hilbert space of a quantum field --- is treated in detail in Appendix \ref{app:continuousproof}.

Writing $\hat U_\textsc{a}$ in the form of Eq.~\eqref{eq:controlled_unitary} allows us to understand the action of $\hat U_\textsc{a}$ as acting with the unitary $\exp\left(-\ii x_k \hat m\right)$ on the target system A, conditional on the source S being in the state $\ket{x_k}$. In this sense, it can even be understood as a measurement of $\hat X$ on the source by the target system. Then, from Eq.~\eqref{eq:ent_breaking_channel}, we see that the final partial state $\rhoa$ of A following its interaction with S is
\begin{align}
    \hat \rho_\textsc{a} &= \sum_k \bra{x_k}\hat \rho_{0,\textsc{s}}\ket{x_k} \exp\left(-\ii x_k \hat m\right)\hat \rho_{\textsc{a},0} \exp\left(\ii x_k \hat m\right).
\end{align}
This is exactly the general form of an entanglement breaking quantum channel from the source's initial state to the target's final state~\cite{horodecki_general_2003,Holevo2005}.

In the context of entanglement extraction this means that if the target system that is the last to interact with the source is coupling through a single simple-generated unitary, i.e., a unitary of the form in Eq.~\eqref{eq:r1generated}, then that target system always ends up in a separable state with its environment, where in this case the environment includes the other target system. Hence, in this general scenario no entanglement can be extracted by the targets A and B from the source S, irrespective of the specific details of the interaction.

\subsection{Entanglement extraction by combining simple generated couplings}\label{sec:3interactions}

Having seen that it is impossible to entangle the targets A and B using only two simple-generated couplings, the question arises whether it is possible to entangle the targets by combining more than two simple-generated interactions?
In the following we show that it is  possible to get the two targets entangled by coupling one of them once and the second one twice to the source through simple-generated interactions, under certain conditions.

Let us denote the unitaries describing the two interactions of target A with the source S by
\begin{align}
\UA{1}&=\exp\left(-\ii \mA1 \otimes \XA1\right), \\ 
\UA{2}&=\exp\left(-\ii \mA2 \otimes \XA2\right),
\end{align}
and the unitary describing the single coupling of target B to S by
\begin{align}
\UB1=\exp\left(-\ii \mB1\otimes \XB1\right).
\end{align}

It follows from the previous section that if the interaction $\UB1$ takes place last, then the targets A and B always end up in a separable space. Hence, A and B can only get entangled if the coupling between B and S, given by the unitary $\UB1$, takes place before at least one of A's couplings.

When target B is coupled to the source first followed by the two couplings of A, the product of the two couplings $\UA1\UA2$ must not yield an entanglement breaking channel from the field to target A, otherwise A and B would once again end up in a separable state. In order for this not to occur it is necessary that the two observables $\XA1$ and $\XA2$ do not commute.

To see this, suppose instead that \mbox{$\comm{\XA1}{\XA2}=0$}. Then, the two observables $\XA1$ and $\XA2$ can be simultaneously diagonalized as
\begin{align}
    \XA1 &=\sum_k a^{(1)}_k \ketbra{x_{\textsc{a},k}}{x_{\textsc{a},k}},\\
    \XA2 &=\sum_k a^{(2)}_k \ketbra{x_{\textsc{a},k}}{x_{\textsc{a},k}}.
\end{align}
Therefore the product $\UA1\UA2$ of the unitaries governing the interactions between A and S can be expressed as
\begin{align}\label{eq:alice_entbreaking_product}
\sum_k \exp\!\left(\!-\ii a^{(1)}_k \mA1\!\right) \exp\!\left(\!-\ii a^{(2)}_k \mA2\!\right) \!\otimes\! \ketbra{x_{\textsc{a},k}}{x_{\textsc{a},k}}\!,
\end{align}
which again has the form of a controlled unitary gate (performing a unitary on the target system conditional on the source system's state) and, therefore, gives rise to an entanglement breaking channel from the source to target A.

This observation, together with the fact that it is necessary to have $\comm{\XB1}{\XA{n}}\neq0$ in order to obtain $\comm{\UB1}{\UA{n}}\neq0$, leads to the conclusion that if more than one of the three commutators $\comm{\XA1}{\XA2}$, $\comm{\XB1}{\XA1}$ and $\comm{\XB1}{\XA2}$ vanish, then, regardless of the order in which they interact with the source S, the targets A and B always end up in a separable state. This is simply because if two of these commutators vanish then it is always possible to rearrange the product of unitaries $\UB1\UA1\UA2$ (or $\UA1\UB1\UA2$) such that it ends with an entanglement breaking coupling from the system to the corresponding target.

\begin{figure}

\centering




\includegraphics[width=0.25\textwidth]{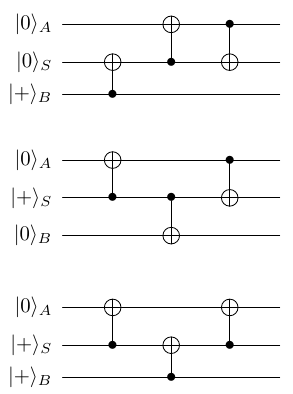}
\caption{Qubit toy models for protocols that entangle the target systems A and B using three CNOT gates with the source S, which are examples of simple-generated interactions. From the top to the bottom the commutators $\comm{\UB1}{\UA2}$, $\comm{\UB1}{\UA1}$ and $\comm{\UA1}{\UA2}$ vanish respectively. The first two examples yield the final state $\frac1{\sqrt2}\left(\ket{00}+\ket{11}\right)_\textsc{ab}\otimes\ket0_\textsc{s}$ whereas the last one yields $\frac1{\sqrt2} \left(\ket{00}+\ket{11}\right)_\textsc{ab} \otimes\ket+_\textsc{s}$.}
  \label{fig:qubittoys3} 
\end{figure}

To demonstrate that entangling two targets via three simple-generated interactions \emph{is} possible if one satisfies the above described necessary condition, we can construct simple toy models where both targets as well the source are modelled by single qubits. In this case we can use the CNOT-gate between two qubits as a simple-generated interaction between target and source,
\begin{align}
\hat U_\textsc{CNOT}&= \exp\left[-\ii\pi \left( 2\ketbra00+\ketbra11\right)\right.\nn
&\left.\qquad\qquad  \otimes \left( 2\ketbra++ + 3\ketbra--\right) \right].
\end{align}
Here, we will allow either the target quantum system or the source quantum system to play the role of the control gate in the CNOT.
Fig.~\ref{fig:qubittoys3} shows examples of circuits that achieve entanglement between the target qubits through an interaction with a single source qubit. In each of the three cases a different commutator $\comm{\UB1}{\UA1}$, $\comm{\UB1}{\UA2}$, or $\comm{\UA1}{\UA2}$ vanishes.

Arguably, the toy models of Fig.~\ref{fig:qubittoys3} do not technically represent entanglement extraction from the source, since the very notion of entanglement extraction from a single qubit onto two qubits does not make sense. Rather, the toy models are showing  a mechanism of entangling the targets through communication via the source.

In fact, the finding above that at most one pair out of $\XB1,\XA1,\XA2$ may commute for entanglement extraction to be possible, implies that three simple-generated interactions can only entangle the target systems if they could alternatively  be used to  implement a communication channel from A to B or vice versa. This is because if a commutator of the form $\comm{\XB1}{\XA1}$ is non-vanishing, then the corresponding pair of interactions could also be used to send information from target A to target B~\cite{dickinson_probabilities_2016,cliche_relativistic_2010}.

There is another  observation which suggests that  entangling two target systems with three simple-generated interactions really corresponds to correlating them through communication rather than extracting entanglement from the source. This is the fact  that it is not possible to genuinely extract entanglement from a source consisting of a pair of qubits with only three simple-generated interactions.

\begin{figure}
\centering

\includegraphics[width=0.25\textwidth]{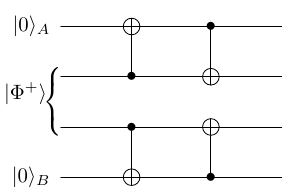}

  \caption{Qubit toy model demonstrating that four simple-generated interactions (here CNOT gates) can extract pre-existing entanglement from a source, which is modeled as a pair of qubits in the maximally entangled state $\ket{\Phi^+}_\textsc{s}=\frac1{\sqrt2}\left(\ket{00}+\ket{11}\right)$. The circuits swaps the entanglement onto the targets. The final state reads $\frac1{\sqrt2}\left(\ket{00}+\ket{11}\right)_\textsc{ab}\otimes\ket{00}_\textsc{s}$.
  }
  \label{fig:toyextraction}
\end{figure}

To see this, we assume that the source is given by a pair of qubits in some entangled state (cf. Fig.~\ref{fig:toyextraction}, which shows genuine extraction from this  toy model using four interactions).
Let B be the target system that couples only once, and hence only interacts with one of the source qubits. Then, in order for A and B to have any chance of extracting pre-existing entanglement from S, the target A needs to use its two interactions to couple to each of the two source qubits once, since otherwise the pre-existing entanglement between the source qubits would not be of any significance to the protocol.

Now, operators that act on only one source qubit commute with operators that act on the other source qubit. This implies that the interaction of B with one source qubit commutes with the interaction of A with the other source qubit. However, both interactions of A with the source also commute with each other because they act on different source qubits. This means that two out of the three possible pairings of observables generating the interactions, $\XB1,\XA1,\XA2$, commute. Thus by the argument above the targets end up in a separable state.

The only possible way to get the two targets to become entangled is to use all three couplings to interact with only one of the source's qubits. Clearly, such a protocol does not access the pre-existing entanglement in the source at all. In fact, entanglement between the accessed source qubit and the other source qubit impedes, rather than facilitates, the entanglement of the two target systems.

In summary, it is possible to achieve entanglement between two target systems with three simple-generated couplings. However, in these  scenarios
the couplings need to be such that they could also be used to send information from one of the targets to the other (not necessarily in both directions): 
The source system seems to play the role of a communication medium which serves to correlate the two targets. 
Genuine extraction of pre-existing entanglement from the source system, e.g., by spacelike separated observers, seems to require at least four simple-generated couplings. A toy model example of this is shown in Fig.~\ref{fig:toyextraction}.

\section{Applications to entanglement harvesting}\label{sec:UDWsection}

A frequently studied physical system to which we will now apply our results is the entanglement harvesting setup, in which two qubits (the targets) attempt to become entangled by interacting with a quantum field (the source). This will allow us, in Sec.~\ref{sec:null_result_ent_harv}, to generalize previous no-go entanglement harvesting results, as well as provide a unified explanation for why they hold. Then in Sec.~\ref{sec:simple_setup_delta_harvesting} we will non-perturbatively explore the simplest coupling scenario between qubits and field (i.e., three Dirac-$\delta$ couplings) through which the qubits are able to harvest field entanglement. In particular, and in contrast to perturbative results, we will show that the amount of extracted entanglement decreases above a certain optimal value for the coupling strength.

To that end, consider the scalar field $\hat \phi(\bm x,t)$ in $(n+1)$-dimensional flat spacetime, which can be expanded in plane-wave modes as
\begin{equation}
\label{eq:field}
	\hat{\phi}(\bm{x},t)
	=
	\int\frac{\d[n]{\bm{k}}}{\sqrt{2(2\pi)^n |\bm{k}|}}\left[\ad{k} e^{\ii(|\bm{k}|t-\bm{k}\cdot\bm{x})}+\text{H.c.}\right].
\end{equation}
We denote the two qubits (also called \textit{detectors}) by $\nu\in\{\text{A},\text{B}\}$, their free ground and excited states by $\ket{g_\nu}$ and $\ket{e_\nu}$, and their energy gaps by $\Omega_\nu$. Suppose that the two detector-field system starts out in the arbitrary separable state
\begin{equation}\label{eq:rho_0}
    \hat \rho = 
    \rhoa\otimes\hat\rho_\phi\otimes\rhob,
\end{equation}
in the Hilbert space $\mathcal H_\textsc{a}\otimes\mathcal H_\phi\otimes\mathcal H_\textsc{b}$. We will assume that the detectors are at rest at positions $\bm x_\nu$ (in the $(\bm x,t)$ coordinate system in which we performed the field quantization), and we let $F_\nu(\bm x)$ be real-valued distributions (dimensions of $L^{-n}$) describing the detectors' spatial profiles. We allow detector $\nu$ to interact with the quantum field through the interaction Hamiltonian (in the interaction picture)
\begin{equation}
\label{eq:H_I_nu}
    \hat{H}_{\textsc{i},\nu}(t)
    =
    \tilde\lambda_\nu \chi_\nu(t) \hat{m}_\nu(t) \otimes
	\!\int \!\d[n]{\bm{x}} F_\nu(\bm{x}-\bm{x}_\nu) \hat{\phi}(\bm{x},t).
\end{equation}
Here $\tilde\lambda_\nu$ is a coupling strength (dimension $L^{(n-3)/2}$), $\chi_\nu(t)$ is a dimensionless switching function that describes how the detector is turned on and off, and $\hat{m}_\nu$ is the monopole moment of detector $\nu$,
\begin{equation}
    \label{eq:m_nu}
	\hat{m}_\nu(t)=
	\ket{e_\nu}\bra{g_\nu} e^{\ii\Omega_\nu t}+
	\ket{g_\nu}\bra{e_\nu} e^{-\ii\Omega_\nu t}.
\end{equation} 
The type of interaction between detector and field given by Eq.~\eqref{eq:H_I_nu} is the well-known Unruh-DeWitt interaction~\cite{DeWitt1979}, which captures the essential features of the light-matter interaction when angular momentum exchange can be ignored~\cite{Martinez2013,Alhambra2014,Pozas2016}.

The result that follows in Sec.~\ref{sec:null_result_ent_harv} can straightforwardly be extended to detectors with arbitrary trajectories, but in this case care must be taken to specify each detector's parameters (energy gap, switching function, smearing function) in the detector's own rest frame, and then perform appropriate coordinate transformations in order to get the interaction Hamiltonian in the lab frame $(\bm x,t)$~\cite{Martinez2018}. In order to avoid going into these details and obscuring our main objective, we will consider only stationary detectors.

\subsection{Null result for entanglement harvesting}\label{sec:null_result_ent_harv}

We will now show examples of two classes of Unruh-DeWitt interactions between two detectors and the field for which the interaction Hamiltonian in Eq.~\eqref{eq:H_I_nu} generates a time-evolution unitary that is of the ``simple" form in Eq.~\eqref{eq:r1generated}, and therefore, by our general result in Sec.~\ref{sec:ent_breaking_channels}, these interactions are unable to extract entanglement from the field to the detectors. These two classes of interactions are i) the case of degenerate detectors~\cite{Braun2002,Braun2005,landulfo_nonperturbative_2016,Pozas2017}, and ii) the case of detectors that couple to the field at one instant in time (i.e., through a Dirac-$\delta$ function)~\cite{Simidzija2017c}. Importantly, both i) and ii) are prevalent interactions considered in the literature, due to their physical significance as well as the fact that they allow for non-perturbative studies of detector-field interactions~\cite{landulfo_nonperturbative_2016,Simidzija2017c}, something that is difficult to do in other regimes.

To that end, suppose now that the interaction Hamiltonian $\hat H_\textsc{i}(t)$ between the detectors and the field, given in Eq.~\eqref{eq:H_I_nu}, is such that the time-evolution unitary $\hat U$ of the system, given in Eqs.~\eqref{eq:U_dyson} and \eqref{eq:U_magnus}, reads
\begin{equation}\label{eq:U=UBUA}
    \hat U = (\mathds{1}_\textsc{a}\otimes\hat U_{\textsc{b}\phi})(\hat U_{\textsc{a}\phi}\otimes\mathds{1}_\textsc{b}),
\end{equation}
where $\hat U_{\nu\phi}$ is a unitary on the Hilbert space $\mathcal H_\nu\otimes\mathcal H_\phi$, and is given 
\begin{align}
    \hat U_{\nu\phi}
    &=
    \mathcal T
    \exp
    \left(
    -\ii
    \int_{-\infty}^{\infty}\dif t'\hat{H}_{\textsc{i},\nu}(t')
    \right).\label{eq:Unuphi}
\end{align}
This form for $\hat U$ is achieved, for instance, if detectors A and B are spacelike separated during the times of their interactions with the field, or, alternatively, if detector A is finished coupling to the field before detector B couples, i.e. if $\text{supp}\,\chi_\textsc{a}(t)\subseteq (-\infty,\tilde t\,]$ and $\text{supp}\,\chi_\textsc{b}(t)\subseteq [\tilde t,\infty)$ for some $\tilde t\in \mathbb{R}$. 

Given the initial separable state $\hat \rho_0$ in Eq.~\eqref{eq:rho_0} for the detectors-field system, and assuming that the system evolves according to the unitary $\hat U$ in Eq.~\eqref{eq:U=UBUA}, we would like to determine whether the two detectors can become entangled (i.e. harvest entanglement) through their interactions with the field. 

We see that the system first evolves according to the unitary $\hat U_{\textsc{a}\phi}\otimes\mathds{1}_\textsc{b}$, which puts it in a state $\hat \rho_{\textsc{a}\phi}\otimes \rhob$, where $\hat \rho_{\textsc{a}\phi}$ is an arbitrary (possibly entangled) state in the Hilbert space $\mathcal H_\textsc{a}\otimes\mathcal{H}_\phi$. Now to obtain the final state of the two detectors, we need to couple Bob to the field by applying the unitary $\mathds 1_\textsc{a}\otimes\hat U_{\textsc{b}\phi}$, and then trace out the field. This corresponds exactly to applying the channel $\mathds 1_\textsc{a}\otimes\xi$ to the state $\hat\rho_{\textsc{a}\phi}$, where $\xi$ maps states in $\mathcal H_\phi$ to states in $\mathcal H_\textsc{b}$ and is defined by
\begin{equation}
    \xi(\hat \rho_\phi):=
    \tr_\phi
    \left[
    \hat U_{\textsc{b}\phi}
    \left(\hat \rho_\phi\otimes\rhob\right)
    \hat U_{\textsc{b}\phi}^\dagger
    \right].
\end{equation}
Now, notice that if the unitary $\hat U_{\textsc{b}\phi}$ is of the form
\begin{equation}\label{eq:Ubphi}
    \hat U_{\textsc{b}\phi}=
    \exp\left(-\ii\hat m_\textsc{b}\otimes\hat X_\textsc{b}\right),
\end{equation}
then from the discussion in Sec.~\ref{sec:ent_breaking_channels} the channel $\xi$ is entanglement breaking, and hence the final state of Alice and Bob is separable. We will now show two example interactions where $\hat U_{\textsc{b}\phi}$ takes the form in Eq.~\eqref{eq:Ubphi}; i.e. interactions for which Alice and Bob cannot harvest entanglement from the field.

\subsubsection{Degenerate detector}

The unitary $\hat U_{\textsc{b}\phi}$ in Eq.~\eqref{eq:Unuphi} can be expressed in the Magnus form \eqref{eq:U_magnus}, i.e., as
\begin{align}\label{eq:U_mag}
    &\hat U =\exp \left(\sum_{n=1}^\infty \hat \Omega_n \right),
\end{align}
where the lowest order terms read
\begin{align}
    &\hat \Omega_1= -\ii \integral{t}{-\infty}\infty \hat H_\textsc{i,b}(t), \label{eq:U_mag1}\\
    &\hat \Omega_2 = -\frac12 \integral{t}{-\infty}\infty \integral{t'}{-\infty}t \comm{\hat H_\textsc{i,b}(t)}{\hat H_\textsc{i,b}(t')},\label{eq:U_mag2}
\end{align}
and the higher-order terms, which are obtained recursively, contain commutators of the commutators of the interaction Hamiltonian $\hat H_\textsc{i,b}(t)$ at increasing orders.

In the case of a degenerate Unruh-DeWitt detector B, i.e. if $\Omega_\textsc{b}=0$, these higher order terms all vanish. (Note that the energy gap $\Omega_\textsc{b}$ of detector B should not be confused with the Magnus expansion terms $\hat \Omega_i$.) To see this, first note from Eq.~\eqref{eq:m_nu} that $\hat m_\textsc{b}$, the tensor factor of $\hat H_\textsc{i,b}(t)$ corresponding to the detector, has no time-dependence if $\Omega_\textsc{b}=0$. Hence
\begin{equation}
    \hat{H}_{\textsc{i,b}}(t)
    =
    \tilde\lambda_\textsc{b} \chi_\textsc{b}(t) \hat{m}_\textsc{b} \otimes\hat\Phi(t),
\end{equation}
where $\hat\Phi(t)$ is the smeared field observable defined by
\begin{equation}
    \hat\Phi(t):=
    \int \!\d[n]{\bm{x}} F_\textsc{b}(\bm{x}-\bm{x}_\nu) \hat{\phi}(\bm{x},t).
\end{equation}
Because the commutator of the field with itself is proportional to the identity, we have that $\comm{\hat H_{\textsc{i,b}}(t)}{\hat H_{\textsc{i,b}}(t')}\propto \hat m_\textsc{b}^2\otimes\id_\phi$, and hence that all higher order commutators of $H_{\textsc{i,b}}$ with itself at different times vanish. Hence $\hat \Omega_k$ is identically zero for all $k\geq3$. Using Eqs.~\eqref{eq:U_mag}, \eqref{eq:U_mag1} and \eqref{eq:U_mag2} then allows us to write $\hat U_{\textsc{b}\phi}$ in the form of Eq.~\eqref{eq:Ubphi}, with $\hat X_\textsc{b}$ defined as 
\begin{align}
    \hat X_\textsc{b}&:=\integral{t}{-\infty}{\infty} \tilde\lambda \chi_\textsc{b}(t) \hat \Phi(t) \\
    &\qquad +\frac\ii2 \integral{t}{\-\infty}{\infty}\integral{t'}{-\infty}{t} \tilde\lambda^2 \chi_\textsc{b}(t)\chi_\textsc{b}(t') \comm{\hat \Phi(t)}{\hat \Phi(t')}\notag.
\end{align}

We therefore arrive at the following conclusion: Suppose that two UDW detectors A and B interact with the field such that i) they are spacelike separated, or ii) detector A interacts with the field strictly before detector B. Then, if detector B is degenerate, the detectors cannot harvest any entanglement from the field. This is a generalization of the perturbative result in~\cite{Pozas2017}, where it was shown that identical, degenerate detectors that satisfy the condition i) or ii), cannot harvest entanglement from the field vacuum. Here, just by investigating the commutator structure of the detector-field interaction Hamiltonian (i.e. without any lengthy calculations), we have shown that this is indeed true in the non-perturbative regime, for non-identical detectors, and for any field state.

\subsubsection{Delta-coupled detector}

Let us now suppose that the switching function for (the possibly non-degenerate) detector B is a delta function,
\begin{equation}
    \chi_\textsc{b}(t)=\eta_\textsc{b}\delta(t-t_\textsc{b}).
\end{equation}
Here $\eta_\textsc{b}$ has dimensions of $L$ and it characterizes the strength of detector B's coupling to the field. Since this is the same as the role played by the coupling strength $\tilde\lambda_\textsc{b}$, we will from here on combine the two as an overall coupling strength $\lambda_\textsc{b}:=\tilde\lambda_\textsc{b}\eta_\textsc{b}$. Hence we are now particularizing our discussion to interactions where detector B interacts with the field at only one instant in time, $t_\textsc{b}$, but with an infinite intensity. Therefore, the total energy exchanged between detector and field is still finite. Such interactions, which we will refer to as $\delta$-couplings, can be viewed as idealized limits of highly intense interactions occurring over short time intervals (see~\cite{Simidzija2017c} for a more detailed discussion). 

Assuming such a switching function, immediately from Eq.~\eqref{eq:Unuphi} we see that the unitary $\hat U_{\textsc{b}\phi}$ governing the time evolution of detector B with the field is of the form in Eq.~\eqref{eq:Ubphi}, with $\hat m_\textsc{b}:=\hat m_\textsc{b}(t_\textsc{b})$ and $\hat X_\textsc{b}$ defined as
\begin{equation}
    \hat{X}_\textsc{b}
    \coloneqq
    \lambda_\textsc{b}
    \int\d[n]{\bm x}
    F_\textsc{b}(\bm{x}-\bm{x}_\textsc{b})
    \hat{\phi}(\bm{x},t_\textsc{b}).
\end{equation}

We therefore arrive at the following conclusion: Suppose that two UDW detectors A and B interact with the field such that i) they are spacelike separated, or ii) detector A interacts with the field strictly before detector B. Then, if detector B interacts with the field at only one instant in time (i.e. through a $\delta$-function, then the detectors cannot harvest any entanglement from the field. This is a generalization of the result obtained in~\cite{Simidzija2017c}, where it was shown that detectors that each couple to the field once cannot harvest entanglement from any coherent state of the field. Here, just by investigating the commutator structure of the detector-field interaction Hamiltonian (i.e. without any lengthy calculations), we have shown that this is indeed true for a much more general class of coupling setups, and for any (not necessarily coherent) state of the field.

\subsection{Simplest setup for entanglement harvesting with \texorpdfstring{$\delta$}{delta}-couplings}\label{sec:simple_setup_delta_harvesting}

The case of Alice and Bob each $\delta$-coupling once to the field has already been studied in Ref.~\cite{Simidzija2017c}. The results of \cite{Simidzija2017c} are a particular example of the general result that we discussed in the previous section: two detectors that each $\delta$-couple to the field once cannot become entangled with one another. In this section we will show the simplest example where the detectors \textit{can} become entangled: Alice (A) coupling twice and Bob (B) once. The three possible ways for this to occur are AAB (A first coupling twice, then B once), ABA, and BAA. As discussed in Sec.~\ref{sec:null_result_ent_harv} we know that the first of these schemes (AAB coupling) is incapable of harvesting entanglement, while the harvesting abilities of detectors in the remaining two coupling setups (ABA and BAA) are constrained by our general findings in Sec.~\ref{sec:3interactions}. 

For simplicity, let us work in $(3+1)$-dimensions and suppose the detectors and field are each in their free ground states, so that the initial state of the system, $\ket{\psi_0}$, reads
\begin{equation}\label{eq:init_state}
    \ket{\psi_0}=
    \ket{g_\textsc{a}}\otimes
    \ket{g_\textsc{b}}\otimes
    \ket{0}
    \in 
    \mathcal H_\textsc{a}\otimes
    \mathcal H_\textsc{b}\otimes
    \mathcal H_\phi.
\end{equation}
Furthermore we suppose the detectors are stationary in the inertial frame in which we performed the field quantization, and that their centers of mass are located at $\bm x_\textsc{a}=\bm x_\textsc{b}=0$. We allow the detectors and field to interact according to the Hamiltonian
\begin{equation}\label{eq:3intHamiltonian}
    \hat{H}_\textsc{i}(t)
    =
    \hat H_{\textsc{i,a}}^{(1)}(t)
    +\hat H_{\textsc{i,a}}^{(2)}(t)
    +\hat H_{\textsc{i,b}}^{(1)}(t),
\end{equation}
where the $\hat H_{\textsc{i},\nu}^{(i)}(t)$ is defined as
\begin{equation}\label{eq:H_Inu^i}
    \hat H_{\textsc{i},\nu}^{(i)}(t)
    =
    \lambda_\nu \delta(t-t_{\nu i}) \hat{m}_\nu(t)\otimes
	\int \d[3]{\bm{x}} F_\nu(\bm{x}) \hat{\phi}(\bm{x},t).
\end{equation}
We will take $\lambda_\textsc{a}=\lambda_\textsc{b}/2=\lambda$ so that detector A (which couples twice to the field) and detector B (which couples once) interact with the field with the same overall ``total strength". The time-evolution unitary $\hat U$ generated by the interaction Hamiltonian \eqref{eq:3intHamiltonian} is given by
\begin{equation}\label{eq:U}
    \hat U =
    \begin{cases}
        \hat U_{{\textsc{b}}_1}
        \hat U_{\textsc{a}_2}
        \hat U_{\textsc{a}_1}
        \quad&\text{if }
        t_{\textsc{a}_1}
        \le t_{\textsc{a}_2}
        \le t_{\textsc{b}_1},\\
        \hat U_{\textsc{a}_2}
        \hat U_{\textsc{b}_1}
        \hat U_{\textsc{a}_1}
        \quad&\text{if }
        t_{\textsc{a}_1}
        \le t_{\textsc{b}_1}
        \le t_{\textsc{a}_2},\\
        \hat U_{\textsc{a}_2}
        \hat U_{\textsc{a}_1}
        \hat U_{\textsc{b}_1}
        \quad&\text{if }
        t_{\textsc{b}_1}
        \le t_{\textsc{a}_1}
        \le t_{\textsc{a}_2},
    \end{cases}
\end{equation}
where $\hat U_{\nu i}$ is the unitary generated by $\hat H_{\textsc{i},\nu}^{(i)}$. 

We will set the detector smearing function $F_\nu(\bm x)$ to be
\begin{equation}\label{eq:F}
    F_\nu(\bm x)=
    \frac{3}{4\pi\sigma^3}
    \Theta\left(1-\frac{|\bm x|}{\sigma}\right),
\end{equation}
where $\Theta$ is the Heaviside theta function, $\sigma$ is the spatial width of the detector, and the prefactor $3/4\pi$ is chosen so that $\int\d[3]{\bm x} F(\bm x)=1$. Notice that the support of $F_\nu$ is the sphere of radius $\sigma$ centered at $\bm x=0$. 
Hence if detectors A and B interact with the field through unitaries $\hat U_\textsc{a}$ and $\hat U_\textsc{b}$ at times $t_\textsc{a}$ and $t_\textsc{b}$, then the detectors are fully timelike separated during their interactions if and only if $|t_\textsc{a}-t_\textsc{b}|>2\sigma$. Note also that in (3+1)D flat spacetime $[\hat\phi(x),\hat\phi(x')]\neq0$ if and only if $x$ and $x'$ are null separated. Therefore for our choice of detector smearing, if $|t_\textsc{a}-t_\textsc{b}|>2\sigma$ then $\hat U_\textsc{a}$ and $\hat U_\textsc{b}$ necessarily commute, and by the results of the previous section they cannot harvest entanglement. We will now show to what extent the detectors \emph{can} get entangled when they \emph{are} able to signal to each other (i.e. when they are not completely timelike nor spacelike separated).

The time-evolved state of the two detectors after their interactions with the field, denoted $\rhoab$, is obtained by applying the unitary $\hat U$ in Eq.~\eqref{eq:U} to the initial state $\ket{\psi_0}$ in Eq.~\eqref{eq:init_state} and tracing out the field, i.e.
\begin{equation}
    \rhoab
    =
    \tr_\phi\left(\hat U \ket{\psi_0}\bra{\psi_0}\hat U^\dagger\right).
\end{equation}
We carefully evaluate this expression in Appendix~\ref{app:rhoab}. In the basis $\{    \ket{g_\textsc{a}}\ket{g_\textsc{b}},
\ket{g_\textsc{a}}\ket{e_\textsc{b}},
\ket{e_\textsc{a}}\ket{g_\textsc{b}},
\ket{e_\textsc{a}}\ket{e_\textsc{b}}\}$, $\rhoab$ reads
\begin{equation}\label{eq:rhoab}
\rhoab=
\begin{pmatrix}
    \rho_{11} & 0 & 0 & \rho_{14} \\
    0 & \rho_{22} & \rho_{23} & 0 \\
    0 & \rho_{32} & \rho_{33} & 0 \\
    \rho_{41} & 0 & 0 & \rho_{44} \\
\end{pmatrix},
\end{equation}
where the entries $\rho_{ij}$ are dependent on the choice of unitary in Eq.~\eqref{eq:U}, and are also evaluated in Appendix~\ref{app:rhoab}.

We will quantify the entanglement of $\rhoab$ using the negativity $\mathcal N$, which is an entanglement monotone that vanishes only for separable states~\cite{Horodecki1996,Peres1996}. The negativity is defined as $\mathcal N:=-\sum_i \min(E_i,0)$, where $E_i$ are the eigenvalues of the partial transpose of $\rhoab$ (with respect to either system A or B). From Eq.~\eqref{eq:rhoab}, we find the $E_i$ to be
\begin{align}
    E_1&=\frac{1}{2}\left(\rho_{22}+\rho_{33}+\sqrt{\left(\rho_{22}-\rho_{33}\right)^2+4|\rho_{23}|^2}\right),\\
    E_2&=\frac{1}{2}\left(\rho_{22}+\rho_{33}-\sqrt{\left(\rho_{22}-\rho_{33}\right)^2+4|\rho_{23}|^2}\right),\\
    E_3&=\frac{1}{2}\left(\rho_{11}+\rho_{44}+\sqrt{\left(\rho_{11}-\rho_{44}\right)^2+4|\rho_{14}|^2}\right),\\
    E_4&=\frac{1}{2}\left(\rho_{11}+\rho_{44}-\sqrt{\left(\rho_{11}-\rho_{44}\right)^2+4|\rho_{14}|^2}\right).
\end{align}

The eigenvalues $E_i$ of the partial transpose of $\rhoab$, and hence the negativity $\mathcal N$, are functions of the following parameters: the times $\ta{1}$, $\ta{2}$, and $\tb{1}$ at which the detectors couple to the field, the strength $\lambda$ with which the detectors couple to the field, as well as the energy gaps $\oa$ and $\ob$ of the detectors. We investigate each of these dependencies below. For simplicity we will set the units of length to be $\sigma$, which is half the spatial width of a detector. Hence the units of energy and $\lambda$ (in 3+1 dimensions) are $\sigma^{-1}$.

First suppose that $\ta{1}\le \ta{2}\le \tb{1}$. With these constraints, we are not able to find any parameter values which give a non-zero negativity. This is, of course, expected from our result in Sec.~\ref{sec:null_result_ent_harv}: the single coupling between detector B and the field is entanglement breaking, and hence if B couples last, regardless of the way A couples the final state of A and B will be separable.

\begin{figure}
    \centering
    \includegraphics[width=0.45\textwidth]{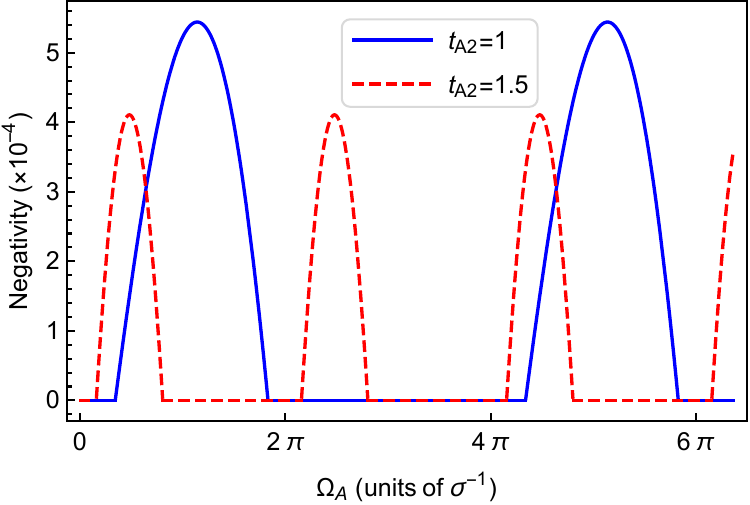}
    \caption{Negativity $\mathcal N$ of a two qubit system as a function of the energy gap $\oa$ of qubit A. Here the coupling scheme is BAA, $\tb{1}=0$, $\ta{1}=0.5$, $\lambda=0.1$, and recall that $\sigma$ is the spatial width of the detectors. The plot is the same for all values of $\ob$ since detector B only couples once. We plot the results for two values of $\ta{2}$. Notice that $\mathcal{N}$ is periodic in $\oa$ with period $2\pi/(\ta{2}-\ta{1})$.}
    \label{fig:neg_vs_omega}
\end{figure}

What happens if we constrain the coupling times by $\tb{1}\le\ta{1}\le\ta{2}$? In this case, we do find parameter values for which the negativity is non-vanishing, as is shown in Fig.~\ref{fig:neg_vs_omega}. In this plot we see that the negativity is a periodic function of the energy gap $\oa$ of detector A, with period $T=2\pi/(\ta{2}-\ta{1})$. This is due to the fact that detector A evolves freely for a time interval $\ta{2}-\ta{1}$ between its two couplings with the field. 
Adding a multiple of $T$ to the detector's free frequency $\oa$ will not alter the phase it picks up during its free evolution. Notice also from  Fig.~\ref{fig:neg_vs_omega} that if the phase difference $\Omega_\textsc{a}(\ta{2}-\ta{1})$ is a multiple of $2\pi$ (i.e. $\Omega_\textsc{a}\in 4\pi\mathbb{Z}$ for the solid curve, and $\Omega_\textsc{a}\in 2\pi\mathbb{Z}$ for the dashed curve), then the detectors cannot harvest entanglement. This comes about because such a phase difference ensures that Alice's two couplings to the field are through the same detector observable, and hence they result in a unitary that is the exponential of a Schmidt rank 1 operator, which, as we have shown, results in an entanglement breaking channel. Reassuringly, we also find the negativity to be independent of $\ob$. This is as expected: since detector B interacts with the field at only one instant in time, any observable phenomenon (like the negativity) is independent of its free evolution, and thus its frequency $\Omega_\textsc{b}$.

These findings allow us to strongly weigh in on the discussion presented in Ref.~\cite{Simidzija2017c}, where the authors found that two detectors that each $\delta$-couple to the field cannot extract any entanglement. Two possible physical explanations were suggested: i) that the sudden $\delta$-couplings induced too much local noise, which is known to have adverse effects on the amount of harvestable entanglement~\cite{Reznik2005,Pozas2015}, or ii) that the lack of harvestable entanglement was a result of each detector not experiencing any free dynamics due to the fact that it only couples to the field at one instant in time. The second explanation is nicely complemented by the perturbative result that degenerate detectors, which also experience a lack of free dynamics, \textit{cannot} harvest entanglement from the vacuum at leading order~\cite{Pozas2017}. We now see that this intuition in ii) seems to be correct. Namely, we have shown that it is indeed possible to harvest entanglement by $\delta$-coupling to the field (therefore the noisy nature of $\delta$-couplings cannot be a critical constraint), but it is necessary for at least one of the detectors to couple more than once to the field (and hence experience non-trivial free evolution). 

\begin{figure}
    \centering
    \includegraphics[width=0.45\textwidth]{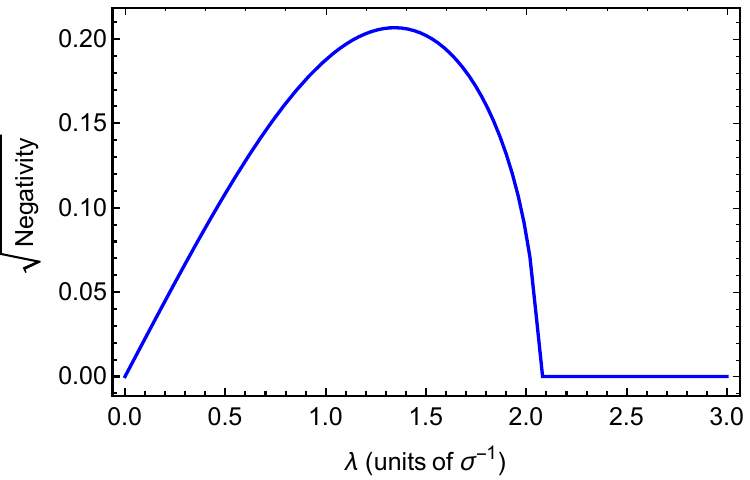}
    \caption{Square root of negativity $\mathcal N$ as a function of the strength $\lambda$ with which detectors couple to the field. Here the coupling scheme is BAA, $\tb{1}=0$, $\ta{1}=0.5$, $\ta{2}=1$, $\oa=3$, and recall that $\sigma$ is the spatial width of the detectors. The plot does is the same for all values of $\ob$ since detector B only couples once. As expected, $\mathcal N\sim \lambda^2$ for $\lambda\ll 1$. Interestingly the dependence is drastically different in the non-perturbative ($\lambda\gtrsim 1$) regime.}
    \label{fig:neg_vs_lambda}
\end{figure}

Let us now explore the dependence of the negativity on the coupling strength $\lambda$ of the detectors to the field. In Fig.~\ref{fig:neg_vs_lambda}, we notice that in the weak-coupling regime $\lambda\ll 1$ (in units of $\sigma^{-1}$), $\mathcal N$ scales as $\lambda^2$. This is a familiar result from perturbative studies~\cite{Pozas2015,Simidzija2017b}, where it has been shown that the leading order contribution to $\mathcal N$ is of $\mathcal{O}(\lambda^2)$. Notice however, that this trend does not continue into the non-perturbative ($\lambda\gtrsim1$) regime. In fact, remarkably, $\mathcal N$ reaches a maximum and then rapidly drops to zero at a finite value of $\lambda$, remaining zero thereafter. That is, in the strong coupling regime, \textit{increasing the coupling strength seems to be detrimental to entanglement harvesting}, at least for delta-couplings. It is possible that this phenomenon is due to the ``noisy" nature of $\delta$-couplings becoming significant in this regime, but more work needs to be done to confirm this.

\begin{figure}
    \centering
    \includegraphics[width=0.45\textwidth]{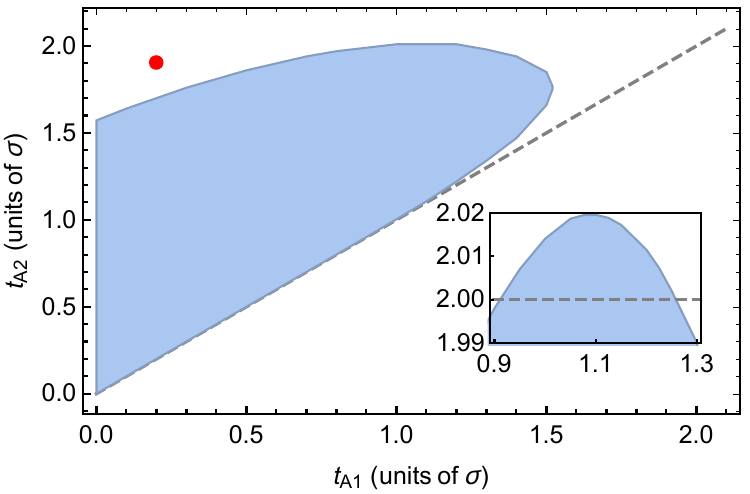}
    \caption{The shaded region indicates values of $\ta{1}$ and $\ta{2}$ for which detectors A and B can become entangled ($\mathcal N>0$) with an appropriate choice of $\oa$.  Note that in the entire shaded region $[\hat U_{\textsc{b}_1},\hat U_{\textsc{a}_1}]$ and $[\hat U_{\textsc{a}_1},\hat U_{\textsc{a}_2}]$ are non-zero, while the inset shows that $\mathcal N>0$ is possible even if $[\hat U_{\textsc{b}_1},\hat U_{\textsc{a}_2}]=0$, i.e., if $\ta{2}>2$. The point at $(\ta{1},\ta{2})=(0.2,1.9)$ shows that $\mathcal N$ could be zero even if none of the three commutators vanish.    Here the coupling scheme is BAA, $\tb{1}=0$, $\lambda\ll 1$ such that $N\sim\lambda^2$, and $\oa$ is arbitrary. The dashed line in the main plot shows $\ta{1}=\ta{2}$.}
    \label{fig:ta1_vs_ta2}
\end{figure}

To conclude this section, let us consider how the times at which the detectors couple to the field affect whether they can become entangled. Concretely, let us again consider the BAA coupling scheme, where we set $\tb{1}=0$. From Fig.~\ref{fig:ta1_vs_ta2}, we see that there is only a finite region in the $\ta{1}-\ta{2}$ plane in which the detectors, by appropriately tuning the energy gap $\oa$, could become entangled. This can be contrasted with the result in~\cite{Pozas2015}, where it was shown that (spacelike separated) detectors with Gaussian switching profiles can always harvest entanglement by increasing their energy gaps, regardless of separation distance.  

Our result for $\delta$-coupled detectors can be understood by our result in Sec.~\ref{sec:3interactions}: in order for two detectors $\delta$-coupling to the field three times in total to become entangled, the values of $\ta{1}$ and $\ta{2}$ (with $\tb{1}$ fixed) must be such that at least two of the three unitary commutators $[\hat U_{\textsc{b}_1},\hat U_{\textsc{a}_1}]$, $[\hat U_{\textsc{b}_1},\hat U_{\textsc{a}_2}]$, and $[\hat U_{\textsc{a}_1},\hat U_{\textsc{a}_2}]$ are non-vanishing. (Recall that in 3+1D flat spacetime, unitaries $\hat U_\textsc{a}$ and $\hat U_\textsc{b}$ at times $t_\textsc{a}$ and $t_\textsc{b}$ commute iff the detectors at these times are not in null contact.) Indeed, the shaded region in Fig.~\ref{fig:ta1_vs_ta2} corresponds to values of $\ta{1}$ and $\ta{2}$ that satisfy this property. We notice however that this commutator condition on entanglement extraction is necessary but not sufficient: there exist values of $\ta{1}$ and $\ta{2}$ (for example $\ta{1}=0.2$, $\ta{2}=1.9$) for which at least two unitary commutators are non-vanishing, yet for which entanglement harvesting is not possible.

\section{Conclusions}\label{sec:conclusions}

We have investigated the reasons why, in the analysis of entanglement harvesting from quantum fields, there were known regimes where entanglement harvesting was not possible. Prompted by these no-go results, we have studied the more general problem of entangling a bipartite separable system through  bi-local interactions with a bipartite entangled source.

Concretely, we have considered the general setup of a separable target system A-B  interacting locally with an entangled source S. Assuming knowledge of the  Hamiltonian governing the time evolution of the system,  we addressed  the pertinent question: under what conditions does the target system A-B become entangled following the interaction?

For a general class of  Hamiltonians $\hat H$ that are frequently considered in the literature, we found a necessary condition that $\hat H$ must obey in order for A and B to be able to extract entanglement from the source. Namely, we showed that if the time-evolution unitary $\hat U$ generated by $\hat H$ is of the form $\hat U = (\mathds{1}_\textsc{a}\otimes\hat U_\textsc{bs})(\hat U_\textsc{as}\otimes\mathds{1}_\textsc{b})$ (i.e target A interacts with the source before target B), and $\hat U_\textsc{b}$ is the exponential of a Schmidt rank 1 operator, then A and B \emph{cannot} become entangled via their interactions with the source.

With this result we have generalized all previously known no-go theorems for entanglement harvesting \cite{Pozas2017,Simidzija2017c}.   The significance of this result arises from the fact that Hamiltonians satisfying the above conditions are commonly used in non-perturbative studies of first quantized systems interacting with quantum fields~\cite{Landulfo2016,Simidzija2017c}. Hence the criterion stated above can be used to  prove non-perturbative results for these systems.

For instance, our general result generalizes one of the main results of Ref.~\cite{Pozas2017}. There it was shown that, to leading order in perturbation theory, identical and degenerate UDW detectors with non-overlapping switching functions cannot harvest any entanglement from the field vacuum in a flat spacetime of any dimensionality. In Ref.~\cite{Pozas2017} it is also shown that for degenerate detectors with overlapping switchings, and spherically symmetric smearings, entanglement harvesting is only possible  in timelike separation. Our result extends this claim to the non-perturbative regime, for not necessarily identical detectors of any shape, and for any arbitrary field state.

There is an important advantage to the method we used to achieve these generalizations here: The conclusions  followed from a direct inspection of the system's Hamiltonian without the need to first  explicitly evaluate the final state of the detectors.

Similarly,  we were able  to generalize the result that two UDW detectors (not necessarily degenerate), each interacting with the quantum field through a single Dirac-$\delta$ coupling, cannot harvest any entanglement from any coherent field state~\cite{Simidzija2017c}. Namely, we find that this is the case for \emph{any} arbitrary field state. Again, with our general criterion at hand, this particular result follows immediately from an inspection of the system's Hamiltonian, without the need for  laborious calculations of the system's time evolution.

Finally, having seen that two $\delta$-couplings are not enough to entangle a pair of UDW detectors, we showed the simplest example of a coupling scheme in which the detectors \emph{do} become entangled   through $\delta$-interactions. For detectors that are able to communicate, i.e. timelike separated, three $\delta$-couplings are sufficient (two for detector A, one for detector B), while for spacelike separated detectors four $\delta$-couplings are required (two per detector).

For the case of three couplings,  we found  a remarkable non-perturbative result:
When the coupling strength $\lambda$ between the detectors and the field is small (compared to other scales with the same dimensions), then the amount of entanglement harvested by the detectors grows as $\lambda^2$.
This was expected from previous perturbative studies~\cite{Pozas2015}.
However, as $\lambda$ exits the perturbative regime this trend reverses itself: the extracted entanglement begins to decrease, and for $\lambda$ larger than some critical value the detectors are not able to extract any entanglement from the field at all.
We conjecture that this is  due to the ``noisy" nature of the sharp and intense $\delta$-couplings, which may manifest itself only in the non-perturbative, strong coupling regime.

The results of this article give rise to new questions in the context of entanglement extraction, in general, and entanglement harvesting from quantum fields, in particular. For instance, our results reveal that it may be necessary to distinguish between the genuine extraction of pre-existing entanglement from a source, and the generation of entanglement between the targets through communication-assisted correlation via the source.
This was illustrated by the qubit toy models which demonstrated how two target systems can become entangled by three simple generated interaction unitaries. There, pre-existing entanglement in the source system was not required, and would in fact be a hindrance to achieving entanglement between the targets.
Furthermore we showed that, in full generality,  when no communication between A and B is possible, i.e., when their couplings to the source commute with each other, then at least two simple generated interactions per target are necessary to achieve genuine entanglement extraction from the source.

It is particularly interesting to apply these considerations to entanglement harvesting from relativistic fields, where the ability to communicate between the targets is determined by their separation being spacelike, null or timelike.
Here, our earlier discussion implies that $\delta$-coupled detectors that are spacelike separated need at least four interactions to extract entanglement from the field. Protocols that only use three $\delta$-couplings in total, meanwhile, can only succeed in extracting entanglement if the detectors are located such that they can communicate via the field.
All these factors together suggest that the triple $\delta$-coupling protocols, while entangling the targets through detector-field coupling, may not be an example of genuine harvesting of entanglement from the field. Instead,  one would need to use at least four $\delta$-coupling to truly harvest pre-existing entanglement from the field's degrees of freedom onto the target detectors.

As a final remark, another direction in which these results could be extended is to investigate how close to a simple generated interaction a target-source interaction can be in order for it to allow for entanglement extraction. It is likely that there is a larger class of interactions, containing the simple generated interactions, for which entanglement extraction still is not possible.
A concept that may be useful to achieve this generalization  is the class of entanglement-annihilating channels~\cite{moravcikova_entanglement-annihilating_2010,filippov_local_2012,filippov_bipartite_2013}. In particular, the entanglement breaking channels that we considered in this article are a strict subset of the 2-locally entanglement-annihilating channels~\cite{moravcikova_entanglement-annihilating_2010}. However, bipartite (or more generally $k$-partite) entanglement extraction is impossible already if  the source-target interaction yields a 2-locally (or generally $k$-locally) entanglement-annihilating quantum channel from the initial state of the source to the final individual partial states of the targets.
\vspace{3mm}

\acknowledgements

The authors would  like to thank Vern Paulsen very much for helpful discussions and insights into the mathematics behind the proof presented in Appendix A. The work of P.S. and E.M.-M. is supported by the Natural Sciences and Engineering Research Council of Canada through the CGS M scholarship and the Discovery program. E.M.-M. also gratefully acknowledges the funding of his Ontario Early Research Award. R.H.J. acknowledges support by ERC Advanced grant 321029 and by the VILLUM FONDEN via the QMATH center of excellence (grant no.10059).

\onecolumngrid
\appendix

\section{Continuous version of controlled unitary argument}\label{app:continuousproof}

Let $X$ be a set, $M$ a $\sigma$-algebra of subsets of $X$, $\mathcal H$ a Hilbert space, and $\hat \mu$ a $\mathcal B(\mathcal H)$-valued measure on $\sigma$. Let $\ket{\phi}\in\mathcal{H}$. Then it is straightforward to show that $\mu_\phi:M\rightarrow \mathbb R_+$ defined by $\mu_\phi(B):=\bra{\phi}\hat\mu(B)\ket{\phi}$ is a positive measure on $M$.

\textbf{Lemma 1: }Let $\hat \rho\in\mathcal{B}(\mathcal{H})$ be a density matrix on $\mathcal H$. Then $\mu_\rho:M\rightarrow \mathbb R_+$ defined by $\mu_\rho(B):=\tr(\mu(B)\hat \rho)$ is a positive measure on $M$.

\textit{Proof: }Write $\hat \rho=\sum_{i=1}^\infty \alpha_i\ket{\phi_i}\bra{\phi_i}$ with $\ket{\phi_i}$ an orthonormal basis, $\alpha_i\ge 0$, and $\sum_{i=0}^\infty\alpha_i<+\infty$. Then 
\begin{equation}\label{eq:mu_rho}
    \mu_\rho(B)
    :=\tr(\mu(B)\hat \rho)
    =\sum_{j=1}^\infty \bra{\phi_j}\hat \mu(B)\hat \rho\ket{\phi_j}
    =
    \sum_{j=1}^\infty \bra{\phi_j}\hat \mu(B) \sum_{i=1}^\infty \alpha_i\ket{\phi_i}\bra{\phi_i} \ket{\phi_j}
    =
    \sum_{i=1}^\infty \alpha_i\bra{\phi_i}\hat \mu(B) \ket{\phi_i}
    =
    \sum_{i=1}^\infty \alpha_i\mu_{\phi_i}(B).
\end{equation}
Hence $\mu_\rho(B)\ge 0$ for all $B\in M$ and $\mu(\varnothing)=0$ since $\alpha_i\ge 0$ and since $\mu_{\phi_i}$ is a positive measure for all $\ket{\phi_i}$. Also
\begin{equation}
    \mu_\rho\left(\bigcup_{k=1}^\infty B_k\right)
    =
    \sum_{i=1}^\infty \alpha_i \mu_{\phi_i}\left(\bigcup_{k=1}^\infty B_k\right)
    =
    \sum_{i=1}^\infty \alpha_i \sum_{k=1}^\infty \mu_{\phi_i}\left(B_k\right), 
\end{equation}
where in the last step we use the fact that $\mu_\phi$ is a measure. Since, additionally, $\mu_\phi$ is positive, we can commute the two summations. Hence
\begin{equation}
    \mu_\rho\left(\bigcup_{k=1}^\infty B_k\right)
    =
    \sum_{k=1}^\infty \sum_{i=1}^\infty \alpha_i \mu_{\phi_i}\left(B_k\right)
    =
    \sum_{k=1}^\infty \mu_\rho\left(B_k\right).
\end{equation}
Hence $\mu_\rho$ is a positive measure. \QED

\textbf{Lemma 2: }Let $f:X\rightarrow \mathbb C$ be a bounded, measurable function. Then $\tr(\int_X f \dif\hat\mu\hat \rho)=\int_X f \dif\mu_\rho$.

\textit{Proof: }Write $\hat \rho=\sum_{i=1}^\infty \alpha_i\ket{\phi_i}\bra{\phi_i}$ with $\ket{\phi_i}$ an orthonormal basis, $\alpha_i\ge 0$, and $\sum_{i=0}^\infty\alpha_i<+\infty$. Then
\begin{equation}
    \tr\left(\int_X f \dif\hat\mu\hat \rho\right)
    =\sum_{i=1}^\infty \alpha_i \bra{\phi_i} \int_X f\dif\hat\mu \ket{\phi_i}
    =\sum_{i=1}^\infty \alpha_i\int_X f\dif \mu_{\phi_i}
    =\int_X f\dif\mu_\rho,
\end{equation}
where in the second equality we used the definition of an operator-valued integral (see, e.g., Ref.~\cite{Hall2013} for details), and in the last equality we made use of Eq.~\eqref{eq:mu_rho}. \QED

Suppose now that $\mathcal H_\textsc{b}$ is a Hilbert space of dimension 2, and $\mathcal H_\phi$ is an infinite dimensional Hilbert space. Note that the argument presented here is straightforwardly extended for any finite value of $\dim \mathcal H_\textsc{b}$. Consider a unitary $\hat U$ on $\mathcal H_\textsc{b}\otimes\mathcal H_\phi$ given by
\begin{equation}\label{eq:U_app}
    \hat U=\exp(-\ii \hat m
    \otimes \hat X),
\end{equation}
where both $\hat m$ and $\hat X$ are self-adjoint operators in their respective Hilbert spaces. Since $\hat m$ is self-adjoint, by the spectral theorem its eigenvalues span $\mathcal H_\textsc{b}$, and in this basis $\hat m$ can be expressed as
\begin{equation}
    \hat m=
    \begin{pmatrix}
        m_{11} & 0\\
        0 & m_{22}
    \end{pmatrix}.
\end{equation}
Since $\hat X$ is self-adjoint, by the spectral theorem for operators on infinite-dimensional Hilbert spaces it can be expressed as (see, e.g.~\cite{Hall2013})
\begin{equation}
    \hat X=
    \int
    \lambda \dif\hat\mu(\lambda),
\end{equation}
where $\hat\mu$ is an operator-valued measure on the Borel $\sigma$-algebra of subsets of the spectrum of $\hat X$. For any measurable function $f$ we define $f(\hat X)$ to be the operator
\begin{equation}\label{eq:f(X)}
    f(\hat X):=
    \int
    f(\lambda) \dif\hat\mu(\lambda).
\end{equation}

Expanding $\hat U$ in Eq.~\eqref{eq:U_app} in a power series gives
\begin{equation}
    \hat U
    =
    \sum_{n=0}^\infty 
    \frac{1}{n!}(-\ii)^n \hat m^n \hat X^n
    =
    \sum_{n=0}^\infty 
    \frac{1}{n!}(-\ii)^n
    \begin{pmatrix}
        m_{11}^n\hat X^n & 0\\
        0 & m_{22}^n\hat X^n
    \end{pmatrix}
    =
    \sum_{n=0}^\infty 
    \frac{1}{n!}(-\ii)^n
    \begin{pmatrix}
        m_{11}^n\int
    \lambda^n \dif\hat\mu(\lambda) & 0\\
        0 & m_{22}^n\hat \int
    \lambda^n \dif\hat\mu(\lambda)
    \end{pmatrix},
\end{equation}
where in the second equality we are representing vectors in $\mathcal H_\textsc{b}$ in the eigenbasis of $\hat m$, and in the third equality we have made use of Eq.~\eqref{eq:f(X)}. By linearity of integration this can be written as
\begin{equation}
    \hat U =
    \begin{pmatrix}
        \int
        \sum\limits_{n=0}^\infty 
    \frac{1}{n!}
    (-\ii m_{11}\lambda)^n \dif\hat\mu(\lambda) & 0\\
        0 & 
        \int
        \sum\limits_{n=0}^\infty 
    \frac{1}{n!}
    (-\ii m_{22}\lambda)^n \dif\hat\mu(\lambda)
    \end{pmatrix}
    =
    \begin{pmatrix}
        \int
        e^{-\ii m_{11} \lambda} \dif\hat\mu(\lambda) & 0\\
        0 & 
        \int
        e^{-\ii m_{22} \lambda} \dif\hat\mu(\lambda)
    \end{pmatrix}.
\end{equation}
Similarly, the adjoint of $\hat U$, denoted $\hat U^\dagger$, can be expressed as
\begin{equation}
    \hat U^\dagger = 
    \begin{pmatrix}
        \int
        e^{\ii m_{11} \lambda} \dif\hat\mu(\lambda) & 0\\
        0 & 
        \int
        e^{\ii m_{22} \lambda} \dif\hat\mu(\lambda)
    \end{pmatrix}.
\end{equation}

Consider now the channel $\xi$ which takes states (density matrices) on $\mathcal H_\phi$ into states on $\mathcal H_\textsc{b}$ and is given by
\begin{equation}
    \xi(\hat \rho_\phi):=
    \tr_\phi\left[
    \hat U \left(
    \rhob\otimes\hat\rho_\phi
    \right) \hat U^\dagger
    \right],
\end{equation}
where $\rhob$ is a density matrix on $\mathcal H_\textsc{b}$. We can represent $\rhob$ in the eigenbasis of $\hat m$ as
\begin{equation}
    \rhob=
    \begin{pmatrix}
        b_{11} & b_{12}\\
        b_{12}^* & b_{22}
    \end{pmatrix}.
\end{equation}
Also in this basis, the expression for $\xi(\hat \rho_\phi)$ takes the form
\begin{align}
    \xi(\hat \rho_\phi)
    &=
    \tr_\phi
    \begin{pmatrix}
        \int
        e^{-\ii m_{11} \lambda} \dif\hat\mu(\lambda) & 0\\
        0 & 
        \int
        e^{-\ii m_{22} \lambda} \dif\hat\mu(\lambda)
    \end{pmatrix}
    \begin{pmatrix}
        b_{11}\hat \rho_\phi & b_{12}\hat \rho_\phi\\
        b_{12}^*\hat \rho_\phi & b_{22}\hat \rho_\phi
    \end{pmatrix}
    \begin{pmatrix}
        \int
        e^{\ii m_{11} \lambda'} \dif\hat\mu(\lambda') & 0\\
        0 & 
        \int
        e^{\ii m_{22} \lambda'} \dif\hat\mu(\lambda')
    \end{pmatrix} \notag\\
    &=
    \tr_\phi
    \begin{pmatrix}
        \int
        e^{-\ii m_{11} \lambda} \dif\hat\mu(\lambda) b_{11}\hat \rho_\phi \int
        e^{\ii m_{11} \lambda'} \dif\hat\mu(\lambda') 
        & 
        \int
        e^{-\ii m_{11} \lambda} \dif\hat\mu(\lambda) b_{12}\hat \rho_\phi \int
        e^{\ii m_{22} \lambda'} \dif\hat\mu(\lambda')\\
        \int
        e^{-\ii m_{22} \lambda} \dif\hat\mu(\lambda) b_{12}^*\hat \rho_\phi \int
        e^{\ii m_{11} \lambda'} \dif\hat\mu(\lambda')
        & 
        \int
        e^{-\ii m_{22} \lambda} \dif\hat\mu(\lambda) b_{22}\hat \rho_\phi \int
        e^{\ii m_{22} \lambda'} \dif\hat\mu(\lambda')
    \end{pmatrix}\notag\\
    &=
    \begin{pmatrix}
        \tr\int
        e^{-\ii m_{11} \lambda} \dif\hat\mu(\lambda) b_{11}\hat \rho_\phi \int
        e^{\ii m_{11} \lambda'} \dif\hat\mu(\lambda') 
        & 
        \tr\int
        e^{-\ii m_{11} \lambda} \dif\hat\mu(\lambda) b_{12}\hat \rho_\phi \int
        e^{\ii m_{22} \lambda'} \dif\hat\mu(\lambda')\\
        \tr\int
        e^{-\ii m_{22} \lambda} \dif\hat\mu(\lambda) b_{12}^*\hat \rho_\phi \int
        e^{\ii m_{11} \lambda'} \dif\hat\mu(\lambda') 
        & 
        \tr\int
        e^{-\ii m_{22} \lambda} \dif\hat\mu(\lambda) b_{22}\hat \rho_\phi \int
        e^{\ii m_{22} \lambda'} \dif\hat\mu(\lambda'
    \end{pmatrix}.\label{eq:xi_app}
\end{align}
Note that integrals with respect to projection-valued measures have the multiplicative property (see, e.g.~\cite{Hall2013})
\begin{equation}
    \int
    f(\lambda) \dif\hat\mu(\lambda)
    \int
    g(\lambda') \dif\hat\mu(\lambda')
    =
    \int
    f(\lambda)g(\lambda) \dif\hat\mu(\lambda).
\end{equation}
Together with the cyclicity of the trace this simplifies Eq.~\eqref{eq:xi_app} to read
\begin{align}
    \xi(\hat \rho_\phi)
    &=t
    \begin{pmatrix}
        \tr\int b_{11} \dif\hat\mu(\lambda)
        \hat \rho_\phi
        & 
        \tr\int b_{12}
        e^{-\ii (m_{11}-m_{22}) \lambda} \dif\hat\mu(\lambda) 
        \hat \rho_\phi
        \vspace{0.1cm}\\
        \tr\int b_{12}^*
        e^{\ii (m_{11}-m_{22}) \lambda} \dif\hat\mu(\lambda)
        \hat \rho_\phi
        & 
        \tr\int b_{22} \dif\hat\mu(\lambda)
        \hat \rho_\phi
    \end{pmatrix}\notag\\
    &=
    \begin{pmatrix}
        \int b_{11} \dif\nu(\lambda)
        & 
        \int b_{12}
        e^{-\ii (m_{11}-m_{22}) \lambda} \dif\nu(\lambda)
        \vspace{0.1cm}\\
        \int b_{12}^*
        e^{\ii (m_{11}-m_{22}) \lambda} \dif\nu(\lambda)
        & 
        \int b_{22} \dif\nu(\lambda)
    \end{pmatrix},\label{eq:xi_app2}
\end{align}
where in the last step we have defined the real-valued measure $\nu$ by $\nu(\cdot):=\tr(\dif\hat\mu(\cdot)\hat\rho_\phi)$ and made use of Lemma 2. Let us now define the $\mathcal B(\mathcal H_\textsc{b})$-valued function $\hat\rho_\textsc{b}(\lambda)$ so that in the eigenbasis of $\hat m$ it reads
\begin{equation}
    \hat\rho_\textsc{b}(\lambda)=
    \begin{pmatrix}
        b_{11} & b_{12}e^{-\ii (m_{11}-m_{22})\lambda}\\
        b_{12}^*e^{\ii (m_{11}-m_{22})\lambda} & b_{22}
    \end{pmatrix}.
\end{equation}
Then Eq.~\eqref{eq:xi_app2} can be written in a basis independent manner as
\begin{equation}
    \xi(\hat \rho_\phi)=
    \int\hat\rho_\textsc{b}(\lambda) \nu(\lambda).
\end{equation}
Finally, by Theorem 2 in~\cite{Holevo2005}, we see that the channel $\xi$ is entanglement breaking.

\section{Calculation of \texorpdfstring{$\hat\rho_{AB}$}{}}\label{app:rhoab}

Here we will show the procedure for calculating the expression for the density matrix $\rhoab$ for each of the three coupling setups that we consider: AAB (first Alice couples twice then Bob once), BAA, and ABA. Notice that the first two scenarios are just limiting cases of the coupling scheme AABB (up to a relabeling of A$\leftrightarrow$B). Similarly the coupling ABA is a limiting case of the four delta-coupling ABBA, where we take the two B couplings to be at the same time. We will work out the details for the AABB coupling, with the calculations for the ABBA setup performed analogously.

Let us therefore consider the case where A and B each delta-couple to the field twice, at times $\ta{1}\le\ta{2}\le\tb{1}\le\tb{2}$, with coupling strengths $\lambda_\textsc{a}=\lambda_\textsc{b}=\lambda/2$. The interaction Hamiltonian is
\begin{equation}\label{eq:4intHamiltonian}
    \hat{H}_\textsc{i}(t)
    =
    \hat H_{\textsc{i,a}}^{(1)}(t)
    +\hat H_{\textsc{i,a}}^{(2)}(t)
    +\hat H_{\textsc{i,b}}^{(1)}(t)
    +\hat H_{\textsc{i,b}}^{(2)}(t),
\end{equation}
with $\hat H_{\textsc{i},\nu}^{(i)}(t)$ defined in Eq.~\eqref{eq:H_Inu^i}. This Hamiltonian generates the time-evolution unitary $\hat U = \hat U_{\textsc{b}_2}\hat U_{\textsc{b}_1}\hat U_{\textsc{a}_2}\hat U_{\textsc{a}_1}$, with the $\hat U_{\nu i}$ given by (see~\cite{Simidzija2017c} for details)
\begin{align}
    \hat U_{\textsc{a}i}
    &=
    \mathds 1_\textsc{a}\otimes
    \mathds 1_\textsc{b}\otimes
    \ca{i}+
    \hat m_{\textsc{a}i}\otimes
    \mathds 1_\textsc{b}\otimes
    \sa{i},\\
    \hat U_{\textsc{b}i}
    &=
    \mathds 1_\textsc{a}\otimes
    \mathds 1_\textsc{b}\otimes
    \cb{i}+
    \mathds 1_\textsc{a}\otimes
    \hat m_{\textsc{b}i}\otimes
    \sb{i},
\end{align}
where $\hat m_{\nu i}:=\hat m_\nu(t_{\nu i})$, $\cnu{i}:=\cosh(\hat Y_{\nu i})$, $\snu{i}:=\sinh(\hat Y_{\nu i})$, and $\hat Y_{\nu i}:=-\ii(\lambda/2)\int\d[n]{\bm x}F_\nu(\bm{x})\hat{\phi}(\bm{x},t_{\nu i})$. The unitary $\hat U$ evolves the initial state $\ket{\psi_0}$ given in Eq.~\eqref{eq:init_state} into the state
\begin{align}
    \hat U\ket{\psi_0}=&
    \ket{g_\textsc{a}}\otimes
    \ket{g_\textsc{b}}\otimes
    \big(
    \cb{2}\cb{1}\ca{2}\ca{1}+
    e^{-\ii\oa(\ta{2}-\ta{1})}\cb{2}\cb{1}\sa{2}\sa{1}\notag\\
    &\hspace{2cm}+
    e^{-\ii\ob(\tb{2}-\tb{1})}\sb{2}\sb{1}\ca{2}\ca{1}+
    e^{-\ii\oa(\ta{2}-\ta{1})}
    e^{-\ii\ob(\tb{2}-\tb{1})}\sb{2}\sb{1}\sa{2}\sa{1}
    \big)\ket{0}
    \notag\\
    +&
    \ket{g_\textsc{a}}\otimes
    \ket{e_\textsc{b}}\otimes
    \big(
    e^{\ii\ob\tb{1}}\cb{2}\sb{1}\ca{2}\ca{1}+
    e^{-\ii\oa(\ta{2}-\ta{1})}
    e^{\ii\ob\tb{1}}\cb{2}\sb{1}\sa{2}\sa{1}\notag\\
    &\hspace{2cm}+
    e^{\ii\ob\tb{2}}\sb{2}\cb{1}\ca{2}\ca{1}+
    e^{-\ii\oa(\ta{2}-\ta{1})}
    e^{\ii\ob\tb{2}}\sb{2}\cb{1}\sa{2}\sa{1}
    \big)\ket{0}
    \notag\\
    +&
    \ket{e_\textsc{a}}\otimes
    \ket{g_\textsc{b}}\otimes
    \big(
    e^{\ii\ob\ta{1}}\cb{2}\cb{1}\ca{2}\sa{1}+
    e^{\ii\oa\ta{2}}\cb{2}\cb{1}\sa{2}\ca{1}\notag\\
    &\hspace{2cm}+
    e^{\ii\ob\ta{1}}e^{-\ii\ob(\tb{2}-\tb{1})}
    \sb{2}\sb{1}\ca{2}\sa{1}+
    e^{\ii\oa\ta{2}}e^{-\ii\ob(\tb{2}-\tb{1})}
    \sb{2}\sb{1}\sa{2}\ca{1}
    \big)\ket{0}
    \notag\\
    +&
    \ket{e_\textsc{a}}\otimes
    \ket{e_\textsc{b}}\otimes
    \big(
    e^{\ii\oa\ta{1}}e^{\ii\ob\tb{1}}
    \cb{2}\sb{1}\ca{2}\sa{1}+
    e^{\ii\oa\ta{2}}e^{\ii\ob\tb{1}}
    \cb{2}\sb{1}\sa{2}\ca{1}\notag\\
    &\hspace{2cm}+
    e^{\ii\oa\ta{1}}e^{\ii\ob\tb{2}}
    \sb{2}\cb{1}\ca{2}\sa{1}+
    e^{\ii\oa\ta{2}}e^{\ii\ob\tb{2}}
    \sb{2}\cb{1}\sa{2}\ca{1}
    \big)\ket{0}.
\end{align}
Using this expression we can calculate the time-evolved density matrix $\rhoab:=\tr_\phi(\hat U\ket{\psi_0}\bra{\psi_0}\hat U^\dagger)$. For example, in the basis $\{\ket{g_\textsc{a}}\ket{g_\textsc{b}},\ket{g_\textsc{a}}\ket{e_\textsc{b}},\ket{e_\textsc{a}}\ket{g_\textsc{b}},\ket{e_\textsc{a}}\ket{e_\textsc{b}}\}$, the (1,1) component of $\rhoab$, denoted $\rho_{11}$, reads
\begin{align}\label{eq:rho11}
    \rho_{11}
    &=
    h(++++++++)+
    h(++++++--)e^{-\ii\oa(\ta{2}-\ta{1})}+
    \notag\\
    &+
    h(++++--++)e^{-\ii\ob(\tb{2}-\tb{1})}+
    h(++++----)e^{-\ii\oa(\ta{2}-\ta{1})}e^{-\ii\ob(\tb{2}-\tb{1})}+
    \notag\\
    &+
    h(--++++++)e^{\ii\oa(\ta{2}-\ta{1})}+
    h(--++++--)+
    \notag\\
    &+
    h(--++--++)e^{\ii\oa(\ta{2}-\ta{1})}e^{-\ii\ob(\tb{2}-\tb{1})}+
    h(--++----)e^{-\ii\ob(\tb{2}-\tb{1})}+
    \notag\\
    &+
    h(++--++++)e^{\ii\ob(\tb{2}-\tb{1})}+
    h(++--++--)e^{-\ii\oa(\ta{2}-\ta{1})}e^{\ii\ob(\tb{2}-\tb{1})}+
    \notag\\
    &+
    h(++----++)+
    h(++------)e^{-\ii\oa(\ta{2}-\ta{1})}+
    \notag\\
    &+
    h(----++++)e^{\ii\oa(\ta{2}-\ta{1})}e^{\ii\ob(\tb{2}-\tb{1})}+
    h(----++--)e^{\ii\ob(\tb{2}-\tb{1})}+
    \notag\\
    &+
    h(------++)e^{\ii\oa(\ta{2}-\ta{1})}+
    h(--------).
\end{align}
Here $h(l_1,l_2,l_3,l_4,l_5,l_6,l_7,l_8):=\bra{0}\hat y_{\textsc{a}_1}^{l_1}\hat y_{\textsc{a}_2}^{l_2}\hat y_{\textsc{b}_1}^{l_3}\hat y_{\textsc{b}_2}^{l_4}\hat y_{\textsc{b}_2}^{l_5}\hat y_{\textsc{b}_1}^{l_6}\hat y_{\textsc{a}_2}^{l_7}\hat y_{\textsc{a}_1}^{l_8}\ket{0}$ for $l_i=\pm1$. In order to evaluate $h$, it is useful write $\hat y_{\nu i}^\pm=[\exp(\hat Y_{\nu i})\pm\exp(-\hat Y_{\nu i})]/2$. The expression for $h$ then becomes
\begin{equation}\label{eq:h}
    h(l_1,l_2,l_3,l_4,l_5,l_6,l_7,l_8)
    =
    \frac{1}{2^8}\sum_{p_j=\pm1}
    \prod_{i=1}^8 f(l_i,p_i) 
    K(p_1,p_2,p_3,p_4,p_5,p_6,p_7,p_8),
\end{equation}
where $K(p_1,p_2,p_3,p_4,p_5,p_6,p_7,p_8):=\bra{0}e^{p_1\ya{1}}e^{p_2\ya{2}}e^{p_3\yb{1}}e^{p_4\yb{2}}e^{p_5\yb{2}}e^{p_6\yb{1}}e^{p_7\ya{2}}e^{p_8\ya{1}}\ket{0}$, and $f(l_i,p_i)$ equals $-1$ if $l_i=p_i=-1$ and $0$ otherwise. 
Next we define the commutators $\ii\theta_\nu\mathds 1_\phi:=[\hat Y_{\nu 2},\hat Y_{\nu 1}]$ and $\ii\theta_{ij}\mathds 1_\phi:=[\hat Y_{\textsc{b}i},\hat Y_{\textsc{a}j}]$, which evaluate to
\begin{equation}
    \theta_\nu
    =
    \ii\int\d[3]{\bm k}\left(\alpha_{\textsc{a}_1}(\bm k)\alpha_{\textsc{a}_2}^*(\bm k)-\text{c.c}\right),\qquad
    \theta_{ij}
    =
    \ii\int\d[3]{\bm k}\left(\alpha_{\textsc{a}j}(\bm k)\alpha_{\textsc{b}i}^*(\bm k)-\text{c.c}\right),
\end{equation}
where $\alpha_{\nu i}(\bm k)$ is defined by
\begin{equation}
    \alpha_{\nu i}(\bm k):=
    -\frac{\ii\lambda}{2\sqrt{2|\bm k|}}
    \tilde F_\nu^*(\bm k)
    e^{\ii|\bm k|t_{\nu i}}.
\end{equation}
$\tilde F_\nu(\bm k)$ is the Fourier transform of the smearing function $F_\nu(\bm x)$, given in Eq.~\eqref{eq:F}. Calculating $\tilde F_\nu(\bm k)$ we obtain
\begin{equation}
    \tilde F_\nu(\bm k)
    :=
    \frac{1}{\sqrt{(2\pi)^3}}
    \int\d[3]{\bm x}F_\nu(\bm x)
    e^{\ii\bm k\cdot\bm x}
    =
    \sqrt{\frac{2}{\pi}}
    \frac{\sin(\sigma|\bm k|)-\sigma|\bm k|\cos(\sigma|\bm k|)}{(\sigma|\bm k|)^3}.
\end{equation}
The expressions for $\theta_\nu$ and $\theta_{ij}$ then work out to be
\begin{align}\label{eq:thetas}
    \theta_\nu&=
    \frac{9\lambda^2}{4\pi^2}
    I_\text{s}(t_{\nu 2}-t_{\nu 1}),\quad
    \theta_{ij}=
    \frac{9\lambda^2}{4\pi^2}
    I_\text{s}(\tb{i}-\ta{j}),\quad\text{where}\\
    I_\text{s}(x)&:=
    \int_0^\infty\!\!\dif k
    \frac{(\sin(k)-k\cos{k})^2}{k^5}\sin(kx)
    =
    \frac{\pi}{96}x(2-|x|)^2(4+|x|)\Theta(2-|x|).
    \label{eq:Is}
\end{align}
Using the Baker-Campbell-Hausdorff formula the expression for $K$ becomes
\begin{align}\label{eq:K}
    K(p_1,p_2,p_3,p_4,p_5,p_6,p_7,p_8)
    &=
    \bra{0}\hat D_\alpha\ket{0}
    \exp
    \Big[-\frac{\ii}{2}\big(
    (p_1-p_8)(p_3+p_6)\theta_{11}+
    (p_2-p_7)(p_3+p_6)\theta_{12}
    \notag\\
    &\hspace{2cm}+
    (p_1-p_8)(p_4+p_5)\theta_{21}+
    (p_2-p_7)(p_4+p_5)\theta_{22}
    \notag\\
    &\hspace{2cm}+
    (p_1-p_8)(p_2+p_7)\theta_\textsc{a}+
    (p_3-p_6)(p_4+p_5)\theta_\textsc{b}
    \big)\Big],
\end{align}
where $\hat D_\alpha$ is the ``displacement operator" of amplitude $\alpha$, given by
\begin{align}
    \hat D_\alpha&:=
    \exp
    \left[
    \int\d[3]{\bm k}
    \left(
    \alpha(\bm k)a_{\bm{k}}^\dagger-
    \alpha(\bm k)^*a_{\bm{k}}
    \right)
    \right],\hspace{1cm}\text{where}\\
    \alpha(\bm k)&:=
    (p_1+p_8)\alpha_{\textsc{a}_1}(\bm k)+
    (p_2+p_7)\alpha_{\textsc{a}_2}(\bm k)+
    (p_3+p_6)\alpha_{\textsc{b}_1}(\bm k)+
    (p_4+p_5)\alpha_{\textsc{b}_2}(\bm k).
    \label{eq:alpha}
\end{align}
$\hat{D}_\alpha$ acts on the vacuum state $\ket{0}$ to create a coherent state of amplitude $\alpha$, which we denote $\ket{\alpha}$. Thus the factor $\bra{0}\hat D_\alpha\ket{0}$ is simply the inner product between $\ket{0}$ (the coherent state of amplitude $0$) and $\ket{\alpha}$. In Appendix A of~\cite{Simidzija2017c} it is shown how to calculate the inner product of two coherent states. The result is
\begin{align}
    \bra{0}\hat D_\alpha\ket{0}
    &=
    \exp\left(-\frac{1}{2}\int\d[3]{\bm k}
    |\alpha(\bm k)|^2\right).
\end{align}
Using the definition of $\alpha(\bm k)$ in Eq.~\eqref{eq:alpha}, this simplifies to
\begin{align}\label{eq:inner_prod}
    \bra{0}\hat D_\alpha\ket{0}
    =
    \exp\Bigg[
    -\frac{9\lambda^2}{16\pi^2}
    \Bigg(&
    \frac{1}{4}\Big(
    (p_1+p_8)^2+
    (p_2+p_7)^2+
    (p_3+p_6)^2+
    (p_4+p_5)^2\Big)
    \notag\\
    &
    +2(p_1+p_8)(p_2+p_7)I_\text{c}(\ta{2}-\ta{1})
    +2(p_1+p_8)(p_3+p_6)I_\text{c}(\tb{1}-\ta{1})
    \notag\\
    &
    +2(p_1+p_8)(p_4+p_5)I_\text{c}(\tb{2}-\ta{1})
    +2(p_2+p_7)(p_3+p_6)I_\text{c}(\tb{1}-\ta{2})
    \notag\\
    &
    +2(p_2+p_7)(p_4+p_5)I_\text{c}(\tb{2}-\ta{2})
    +2(p_3+p_6)(p_4+p_5)I_\text{c}(\tb{2}-\tb{1})
    \Bigg{]},
\end{align}
where $I_\text{c}(x)$ is given by
\begin{align}
     I_\text{c}(x)&:=
    \int_0^\infty\!\!\dif k
    \frac{(\sin(k)-k\cos{k})^2}{k^5}\cos(kx)\notag
    \\
    &=
    \begin{cases}
    \frac{1}{4} &\text{ if }x=0,\\
    \frac{1}{12}(5-8\ln2) &\text{ if }x=\pm2,\\
    \frac{1}{96}
    \Big[
    24+4x^2-2x^2(x^2-12)\ln|x|-16|x|\ln(2+|x|)-12x^2\ln(2+|x|) &\\
    \hspace{1cm}+x^4\ln(2+|x|)+|x|(|x|-2)^2(4+|x|)\ln||x|-2| 
    \Big] &\text{ otherwise.}
    \end{cases}
    \label{eq:Ic}
\end{align}
Substituting Eqs.~\eqref{eq:thetas} and \eqref{eq:inner_prod} into Eq.~\eqref{eq:K} gives us an expression for $K$, which we can then substitute into Eq.~\eqref{eq:h} to get a concrete expression for $h$. Therefore we can get an expression for the matrix element $\rho_{11}$, which is expressed in terms of $h$ in Eq.~\eqref{eq:rho11}. The remaining elements of the two detector density matrix $\rhoab$ are calculated analogously. From the symmetries of the arguments of $h$ in Eq.~\eqref{eq:h} and of $K$ in Eq.~\eqref{eq:K}, we can see that if $h$ has an odd number of ``$-$" arguments then it vanishes. This is the reason why half of the matrix elements of $\rhoab$ in Eq.~\eqref{eq:rhoab} are zero.

\twocolumngrid

\bibliography{references,robs_references}
\bibliographystyle{apsrev4-1}
	
\end{document}